\documentclass[aps,prd,amsfonts,groupedaddress,showpacs,showkeys]{revtex4}
\usepackage{epsfig}
\newcommand{\eq}{\begin{eqnarray}}
\newcommand{\en}{\end{eqnarray}}
\newcommand{\bea}{\begin{eqnarray}}
\newcommand{\eea}{\end{eqnarray}}

\newcommand{\ra}{\rangle}

\begin{document}

\title{
Strong and radiative decays of the $D_{s0}^\ast(2317)$
meson in the  $DK$-molecule picture}
\author{
Amand Faessler,
Thomas Gutsche,
Valery E. Lyubovitskij
\footnote{On leave of absence from the
Department of Physics, Tomsk State University,
634050 Tomsk, Russia},
Yong-Liang Ma
\vspace*{1.2\baselineskip}}

\affiliation{Institut f\"ur Theoretische Physik,
Universit\"at T\"ubingen,
\\ Auf der Morgenstelle 14, D-72076 T\"ubingen, Germany
\vspace*{0.3\baselineskip}\\}

\date{\today}

\begin{abstract}
We consider a possible interpretation of the new charm-strange
meson $D_{s0}^\ast(2317)$ as a hadronic molecule - a bound state
of $D$ and $K$ mesons. Using an effective Lagrangian approach
we calculate the strong $D_{s0}^{\ast} \to D_s \pi^0$ and
radiative $D_{s0}^{\ast} \to D_s^{\ast} \gamma$
decays. A new impact related to the $DK$ molecular structure
of the $D_{s0}^\ast(2317)$ meson is that the presence of $u(d)$
quarks in the $D$ and $K$ mesons gives rise to a direct strong
isospin-violating transition $D_{s0}^{\ast} \to D_s \pi^0$ in
addition to the decay mechanism induced by $\eta-\pi^0$ mixing
considered previously. We show that the direct transition
dominates over the $\eta-\pi^0$ mixing transition in the
$D_{s0}^{\ast} \to D_s \pi^0$ decay.  
Our results for the partial decay widths are consistent  
with previous calculations. 

\end{abstract}

\pacs{13.25.Ft,13.40.Hq,14.40.Lb,14.65.Dw}

\keywords{charm mesons, hadronic molecule, strong and radiative decay,
isospin violation}

\maketitle

\newpage

\section{Introduction}

The complexity of the hadronic mass spectra induces the
possibility that existing and newly observed hadrons can
possibly be interpreted as molecular states (or hadronic
molecules). Such an interpretation is possible, when
the mass of the hadronic molecule $m_H$ lies slightly below
the threshold of the corresponding hadronic pair $H_1 H_2$:
$m_H < m_{H_1} + m_{H_2}$ (for review see e.g.
Refs.~\cite{Voloshin:1976ap}-\cite{Rosner:2006vc}).
In the light meson sector, possible candidates for
hadronic molecules are the scalar mesons $a_0(980)$ and
$\delta(980)$ treated as $K \bar K$ bound
states~\cite{Weinstein:1982gc,Barnes:1985cy,Baru:2003qq}. 
Including the heavy flavor meson sector other possible
molecular states can arise. For example, the scalar and
axial charm $D_{s0}^\ast(2317)$, $D_{s1}(2460)$ and bottom
$B_{s0}^\ast(5725)$ and $B_{s1}(5778)$ mesons
can be  treated as $DK$, $D^\ast K$, $BK$ and $B^\ast K$
bound states~\cite{Barnes:2003dj,Rosner:2006vc,%
vanBeveren:2003kd,Guo:2006fu,Guo:2006rp}, respectively.
Other candidates for a hadronic molecule interpretation are
the $X(3872)$ as a $D^0 \bar D^{\ast 0}$ + charge conjugate (c.c)
bound state, $Y(4260)$ as a $D \bar D_1$ - c.c. and $\psi(4415)$
as a $D_s^\ast \bar D_{s0}(2317)$ + c.c. bound
state~\cite{Rosner:2006vc,Barnes:2005pb}.
In the baryonic sector, the most popular candidate for a hadronic
molecule is the negative-parity $1/2^-$ resonance $\Lambda(1405)$
considered as a $N \bar K$ bound state~\cite{Rosner:2006vc}.
Also, there are candidates in the heavy baryon sector, e.g. the  
charmed baryon $\Lambda_c(2940)^+$ recently discovered by the BABAR  
Collaboration~\cite{Aubert:2006sp} which can be treated as 
a $D^{\ast 0} p$ bound state~\cite{He:2006is}. 

In the current manuscript we focus on the scalar charm-strange meson
$D_{s0}^\ast(2317)$, which was discovered just a few years ago by the
BABAR Collaboration at SLAC in the inclusive $D_s^+ \pi^0$
invariant mass distribution of $e^+ e^-$ annihilation
data~\cite{Aubert:2003fg}. The nearby state $D_{s1}(2460)$ with
a mass of 2.4589 GeV decaying into $D_s^\ast \pi^0$ was observed
by the CLEO Collaboration at CESR~\cite{Besson:2003cp}. Both of these
states have been confirmed by the Belle Collaboration
at KEKB~\cite{Abe:2003jk}.
From interpretation of these experiments it was suggested
that the $D_{s0}^\ast(2317)$ and $D_{s1}(2460)$ mesons are the $P$-wave
charm-strange quark states with spin-parity quantum numbers
$J^P = 0^+$ and $J^P = 1^+$ states, respectively. In the following
the Belle~\cite{Krokovny:2003zq} and the BABAR~\cite{Aubert:2004pw}
Collaborations observed the production of $D_{s0}^\ast(2317)$ and
$D_{s1}(2460)$ in nonleptonic two-body $B$ decays together with
their subsequent strong and radiative transitions. Taking into account
existing experimental information on the  properties of
$D_{s0}^\ast(2317)$ and $D_{s1}(2460)$ mesons~\cite{Yao:2006px}, 
one can conclude that the respective $J^P = 0^+$ and $J^P = 1^+$
quantum numbers are now established with high confidence.

The next important question concerns the possible structure of the
$D_{s0}^\ast(2317)$ and $D_{s1}(2460)$ mesons. The simplest
interpretation of these states is that they are the missing $j_s = 1/2$
(the angular momentum of the $s$-quark) members of the $c \bar s$ $L=1$
multiplet. However, this standard quark model scenario is in
disagreement with experimental observation since the $D_{s0}^\ast(2317)$
and $D_{s1}(2460)$ states are narrower and their masses are lower when
compared to theoretical (see e.g. discussion in Ref.~\cite{Rosner:2006vc}). 
Therefore, in addition to the standard quark-antiquark picture alternative
interpretation of the $D_{s0}^\ast(2317)$ and $D_{s1}(2460)$ mesons have 
been suggested: four-quark states, mixing of two- and four-quark states,
two-diquark states and two-meson molecular states. Up to now different
properties of the $D_{s0}^\ast(2317)$ and $D_{s1}(2460)$ mesons
(masses, strong, radiative and weak decay constants and widths)
have been calculated using different approaches~\cite{Barnes:2003dj},%
\cite{vanBeveren:2003kd}-\cite{Guo:2006rp},%
\cite{Godfrey:2003kg}-\cite{Zhao:2006at}:
quark models, effective Lagrangian approaches,
QCD sum rules, lattice QCD, etc.

In present paper we will consider the strong $D_{s0}^\ast \to D_s + \pi^0$
and radiative $D_{s0}^\ast \to D_s^{\ast} + \gamma$ decays of the
$D_{s0}^\ast(2317)$ meson using an effective Lagrangian approach.
The approach is based on the hypothesis that the $D_{s0}^\ast$
is a strong bound state of $D$ and $K$ mesons. In other words we
investigate the position that $D_{s0}^\ast$ meson is a $(DK)$
hadronic molecule. The coupling of the $D_{s0}^\ast$ meson to the
constituents ($D$ and $K$ mesons) is described by the
effective Lagrangian. The corresponding coupling constant
$g_{D_{s0}^\ast DK}$ is determined by the compositeness
condition $Z=0$~\cite{Weinberg:1962hj,Efimov:1993ei,Lurie:1964qi},
which implies that the renormalization constant of the hadron
wave function is set equal to zero.
Note, that this condition was originally applied to the study of
the deuteron as bound state of proton and neutron~\cite{Weinberg:1962hj}.
Then it was extensively used in the low-energy hadron
phenomenology as the master equation for the treatment of
mesons and baryons as bound states of light and heavy
constituent quarks (see Refs.~\cite{Efimov:1993ei,Efimov:1987na,%
Faessler:2003yf,Efimov:1995uz,Anikin:2000rq}).
In addition this condition was used in Ref.~\cite{Burdanov:2000rw} in
the application to glueballs as bound states of gluons.
Recently the compositeness condition was used to study
the light scalar mesons $a_0$ and $f_0$ as $K \bar K$
molecules~\cite{Baru:2003qq}. 
A new impact of the $DK$ molecular structure of the $D_{s0}^\ast(2317)$
meson is that the presence of $u(d)$ quarks in the $D$ and $K$ meson
gives rise to a direct strong isospin-violating
transition $D_{s0}^{\ast} \to D_s \pi^0$ in addition to
the decay induced by $\eta-\pi^0$ mixing considered before in the
literature. We show that the direct transition dominates over the
$\eta-\pi^0$ mixing transitions. The obtained results for the partial
decay widths are consistent with previous calculations.
By analogy one can treat the second charm narrow resonance $D_{s1}(2460)$
as a $(D^\ast K)$ molecule and the possible corresponding bottom
counterparts - the states $B_{s0}^\ast(5725)$ and $B_{s1}(5778)$ -
as $B K$ and $B^\ast K$ bound states, respectively. The calculation of
the properties of the $D_{s1}(2460)$, $B_{s0}^\ast(5725)$ and
$B_{s1}(5778)$ mesons goes beyond the scope of the present paper and
we relegate this issue to a forthcoming paper. Also in near future
we plan to consider two-body $B$-meson decays and semileptonic processes
involving $D_{s0}^\ast(2317)$ and $D_{s1}(2460)$ in the final state.

In the present manuscript we proceed as follows. First, in Section II
we discuss the basic notions of our approach. We derive the effective
mesonic Lagrangian for the treatment of charm and bottom mesons
$D_{s0}^\ast(2317)$, $D_{s1}(2460)$, $B_{s0}^\ast(5725)$ and
$B_{s1}(5778)$ as $DK$, $D^\ast K$, $BK$ and $B^\ast K$ bound states,
respectively. We discuss how to determine the corresponding coupling
constant between the hadronic molecule and its constituents using
the compositeness condition. In Section III we consider the matrix
elements (Feynman diagrams) describing the strong and radiative decays of
the $D_{s0}^\ast(2317)$. We indicate our numerical results and discuss
various limits, such as the local case and the heavy quark limit.
In Section IV we present a short summary of our results.

\section{Approach}

\subsection{Molecular structure of the $D_{s0}^{\ast \, \pm}(2317)$ meson}

In this section we derive the formalism for the study of the
$D_{s0}^{\ast \, \pm}(2317)$ meson as a hadronic molecule -
a bound state of $D$ and $K$ mesons. First of all we specify
the quantum numbers of the $D_{s0}^{\ast \, \pm}(2317)$ mesons.
We use the current results for the quantum numbers
of isospin, spin and parity: $I(J^P) = 0(0^+)$ and mass
$m_{D_{s0}^{\ast}} = 2.3173$ GeV~\cite{Yao:2006px}.
Our framework is based on an effective interaction
Lagrangian describing the coupling between the $D_{s0}^\ast(2317)$
meson and their constituents - $D$ and $K$ mesons:
\eq\label{Lagr_Ds0}
{\cal L}_{D_{s0}^\ast}(x) \, = \, g_{_{D_{s0}^\ast}} \,
D_{s0}^{\ast \, -}(x) \, \int\! dy \,
\Phi_{D_{s0}^\ast}(y^2) \, D^T(x+w_K y) \, K(x-w_D y) \, + \, {\rm H.c.}
\en
The doublets of $D$ and $K$ mesons are defined as
\eq
D =
\left(
\begin{array}{c}
D^0 \\
D^+ \\
\end{array}
\right)\,, \hspace*{1cm}
K =
\left(
\begin{array}{c}
K^+ \\
K^0 \\
\end{array}
\right)\,,
\en
the symbol $T$ refers to the transpose of the doublet $D$. In particular,
the assumed molecular structure of $D_{s0}^{\ast +}$ and
$D_{s0}^{\ast -}$ states is:
\eq
|D_{s0}^{\ast +}\ra \, = \, |D^+ K^0\ra + |D^0 K^+\ra \,, \hspace*{1cm}
|D_{s0}^{\ast -}\ra \, = \, |D^- \bar K^0\ra + |\bar D^0 K^-\ra \,.
\en
The correlation function $\Phi_{D_{s0}^\ast}$ characterizes the finite
size of the $D_{s0}^\ast(2317)$ meson as a $(DK)$ bound state and
depends on the relative Jacobi coordinate $y$ with
$x$ being the center of mass (CM) coordinate.
Note, the local limit corresponds to the substitution of
$\Phi_{D_{s0}^\ast}$ by the Dirac delta-function:
$\Phi_{D_{s0}^\ast}(y^2) \to \delta^4 (y)$.
The kinematical variables $w_D$ and $w_K$ are defined by
\eq
w_D = \frac{m_D}{m_D + m_K}\,, \hspace*{1cm}
w_K = \frac{m_K}{m_D + m_K}\,, 
\en
where $m_D$ and $m_K$ are the masses of $D$ and $K$ mesons.
The Fourier transform of the correlation function
reads
\eq
\Phi_{D_{s0}^\ast}(y^2) \, = \,
\int\!\frac{d^4p}{(2\pi)^4}  \,
e^{-ip y} \, {\widetilde{\Phi}}_{D_{s0}^\ast}(-p^2) \,.
\en
Any choice for $\tilde\Phi_{D_{s0}^\ast}$ is appropriate
as long as it falls off sufficiently fast in the ultraviolet region
of Euclidean space to render the Feynman diagrams ultraviolet finite.
We employ the Gaussian form
\eq\label{Gauss_CF}
\tilde\Phi_{D_{s0}^\ast}(p_E^2) 
\doteq \exp( - p_E^2/\Lambda^2_{D_{s0}^\ast})\,,
\en
for the vertex function, where $p_{E}$ is the
Euclidean Jacobi momentum. Here $\Lambda_{D_{s0}^\ast}$
is a size parameter, which parametrizes the distribution of
$D$ and $K$ mesons inside the $D_{s0}^\ast$ molecule.

The $D_{s0}^\ast DK$ coupling constant $g_{D_{s0}^\ast}$
is determined by the compositeness 
condition~\cite{Weinberg:1962hj,Efimov:1993ei,Lurie:1964qi},
which implies that the renormalization constant of the hadron
wave function is set equal to zero:
\eq
Z_{D_{s0}^\ast} = 1 -
\Sigma^\prime_{D_{s0}^\ast}(m_{D_{s0}^\ast}^2) = 0 \,,
\en
where
$\Sigma^\prime_{D_{s0}^\ast}(m_{D_{s0}^\ast}^2) =
g_{_{D_{s0}^\ast}}^2 \Pi^\prime_{D_{s0}^\ast}(m_{D_{s0}^\ast}^2)$
is the derivative of the $D_{s0}^\ast$ meson mass operator
described by the diagram in Fig.1.

As we already stressed in Introduction, this condition was
originally applied to the study of the deuteron as a bound state
of proton and neutron~\cite{Weinberg:1962hj}. 
Then it was extensively used in low-energy hadron phenomenology
as the master equation for the treatment of mesons and baryons as
bound states of light and heavy constituent quarks
(see Refs.~\cite{Efimov:1993ei,Efimov:1987na,Faessler:2003yf,%
Efimov:1995uz,Anikin:2000rq}).
In Ref.~\cite{Burdanov:2000rw} this condition was used in the
consideration of glueballs as bound states of gluons.
Recently the compositeness condition was applied to the study
of the light scalar mesons $a_0$ and $f_0$ as $K \bar K$
molecules~\cite{Baru:2003qq}. To clarify the physical
meaning of this condition, we first want to remind the reader that
the renormalization constant $Z_{D_{s0}^\ast}^{1/2}$ can also be
interpreted as the matrix element between the physical and the corresponding
bare state. For $Z_{D_{s0}^\ast}=0$ it then follows that the physical state
does not contain the bare one and is solely described as a bound state.
The interaction Lagrangian Eq.~(\ref{Lagr_Ds0}) and the
corresponding free parts describe both the constituents ($D$ and $K$ mesons)
and the hadronic molecule ($D_{s0}^\ast$), which is taken to be the
bound state of the constituents. As a result of the interaction
the physical particle is dressed, {\it i.e.} its mass and its wave
function have to be renormalized. The condition $Z_{D_{s0}^\ast}=0$
also guarantees that there is no double counting for the physical
observable under consideration: the $D_{s0}^\ast$ meson interacts with
other hadrons and gauge bosons only via its constituents. In particular,
the compositeness condition excludes the direct interaction of the
dressed charged particle (like $D_{s0}^{\ast \, \pm}$ mesons) with the
electromagnetic field. Taking into account both the tree-level
diagram and the diagrams with the self-energy and counter-term
insertions into the external legs (that is the tree-level diagram
times $(Z_{D_{s0}^\ast} - 1))$ one obtains a common factor
$Z_{D_{s0}^\ast}$ which is equal to
zero~\cite{Efimov:1993ei,Efimov:1987na,Faessler:2003yf}.

\subsection{Effective Lagrangian for strong and radiative
decays of $D_{s0}^{\ast \, \pm}(2317)$}

Now we turn to the discussion of the lowest-order diagrams which
contribute to the matrix elements of the strong isospin-violating
decay $D_{s0}^{\ast} \to D_s \pi^0$ and the radiative decay
$D_{s0}^{\ast} \to D_s^\ast \gamma$.
To the strong decay two types of diagrams contribute:
the so-called ``direct'' diagrams of Fig.2 with $\pi^0$-meson
emission from the $D^{(\ast)}K$ meson loops and the ``indirect''
diagrams of Fig.3 where a $\pi^0$ meson is produced via $\eta-\pi^0$
mixing. Note, that the second mechanism based on $\eta-\pi^0$ mixing
was mainly considered before in the literature. Originally,
it was initiated by the analysis based on the use of chiral
Lagrangians~\cite{Cho:1994zu,Bardeen:2003kt,Colangelo:2003vg}
where the leading-order, tree-level $D_{s0}^\ast D_s \pi^0$ coupling
can be generated only by virtual $\eta$-meson emission.
During the last years different approaches have been applied to the
$D_{s0}^{\ast} \to D_s \pi^0$ decay properties using the $\eta-\pi^0$
mixing mechanism. In our approach the $D_{s0}^\ast$ meson
is considered as a $DK$ bound state and, therefore, we have
an additional mechanism for generating the $D_{s0}^\ast D_s \pi^0$
transition due to the direct coupling of $D^{(\ast)}$ and $K^{(\ast)}$
mesons to $\pi^0$. In particular, in the isospin limit (when the masses
of the virtual $D^{(\ast)}$ and $K^{(\ast)}$ mesons in the loops are
degenerate, respectively) the pairs of diagrams related to 
Fig.2(a), 2(b) and Fig.2(c) and 2(d) compensate each other. 
Only the use of physical masses for the $D^{(\ast)}$ and $K^{(\ast)}$ 
mesons gives a nontrivial contribution to the  
$D_{s0}^{\ast} \to D_s \pi^0$ coupling of order $O(\delta)$, where 
\eq\label{delta_par}
\delta \sim m_{D^{(\ast) \, \pm}}^2 - m_{D^{(\ast) \, 0}}^2 \sim
m_{K^{(\ast) \, \pm}}^2 - m_{K^{(\ast) \, 0}}^2
\en
is the parameter of isospin breaking.
Therefore, the contribution of the diagrams of Fig.2 is of the
same order as the one related to Fig.3 involving $\eta-\pi^0$ mixing, 
where the $\eta-\pi^0$ transition coupling (filled black circle) is
counted as $O(\delta)$.

The diagrams contributing to the radiative decay
$D_{s0}^{\ast \, +} \to D_s^{\ast \, +} \gamma$ are shown in Fig.4.
The diagrams of Figs.4(a) and 4(b) are generated by the direct coupling
of the charged $D^+$ and $K^+$ mesons to the electromagnetic field
after gauging of the free Lagrangians related to these mesons.
The diagrams of Figs.4(c) and 4(d) (so-called contact diagrams) are generated
after gauging of nonlocal strong Lagrangian~(\ref{Lagr_Ds0})
describing the coupling of $D_{s0}^{\ast}$ mesons to its constituents -
$D$ and $K$ mesons. The diagrams of Figs.4(e) and 4(f) arise after gauging
the strong $D^\ast_s DK$ interaction Lagrangian containing
derivatives acting on the pseudoscalar fields. Finally, the diagrams of
Figs.4(g) and 4(h) describe the sub-process where the $D_{s0}^\ast$ converts
into the $D_s^\ast$ via a $DK$ loop followed by the interaction
of the $D_s^\ast$ with the electromagnetic field.
Note that an analogous diagram where the $D_s^\ast$ meson
interacts with the electromagnetic field and then converts into the
$D_s^\ast$ vanishes due to the transversity condition for the on-shell
vector meson $D_s^\ast$, i.e. $p_\mu \, \epsilon_{D_s^\ast}^\mu(p) = 0$.
Details of how to generate the effective couplings of the involved
mesons to the electromagnetic field will be discussed later.
 
After the preliminary discussion of the relevant diagrams, now we are
in the position to write down the full effective Lagrangian
${\cal L}_{\rm eff}$ for the study of strong $D_{s0}^{\ast} \to D_s \pi^0$
and radiative $D_{s0}^{\ast} \to D_s^\ast \gamma$ decay properties.
For convenience we split ${\cal L}_{\rm eff}$ into an isospin-symmetric part
${\cal L}_{\rm inv}$ and an isospin-symmetry breaking part
${\cal L}_{\rm break}$:
\eq
{\cal L}_{\rm eff}(x) \, = \,
{\cal L}_{\rm inv}(x) \, + \, {\cal L}_{\rm break}(x) \,,
\en
where ${\cal L}_{\rm inv}$ is given by a sum of free meson parts
${\cal L}_{\rm free}$ and the interaction parts ${\cal L}_{\rm int}$:
\eq
{\cal L}_{\rm inv}(x) \, = \, {\cal L}_{\rm free}(x) \, + \,
{\cal L}_{\rm int}(x) \,.
\en
We use the standard free meson Lagrangian involving states
with quantum numbers $J^P = 0^+, 0^-$ and $1^-$:
\eq
{\cal L}_{\rm free}(x) \, = \,
\sum\limits_{i=S,P,V}  {\cal L}_{\rm free}^i(x) \,,
\en
where
\eq
{\cal L}_{\rm free}^{S}(x) &=&  - D_{s0}^{\ast \, +}(x)
( \Box + m_{D_{s0}^\ast}^2 ) D_{s0}^{\ast \, -}(x) \,,
\label{free0+} \\[2mm]
{\cal L}_{\rm free}^{P}(x) &=&
 - \frac{1}{2} \vec\pi(x) ( \Box + m_\pi^2 ) \vec{\pi}(x)
 - K^\dagger(x) ( \Box + m_K^2 ) K(x)
 - \frac{1}{2} \eta(x) ( \Box + m_\eta^2 ) \eta(x) \nonumber\\[1mm]
 &-& D^\dagger(x) ( \Box + m_D^2 ) D(x)
  - D^+_s(x) ( \Box + m_{D_s}^2 ) D^-_s(x)\,,  \label{free0-}\\[2mm]
{\cal L}_{\rm free}^{V}(x) &=&
  K^{\ast \, \dagger}_\mu (x)
( g^{\mu\nu} [\Box + m_{K^\ast}^2] -
  \partial^\mu\partial^\nu ) K^\ast_\nu(x)
+ D^{\ast \, \dagger}_\mu (x)
( g^{\mu\nu} [\Box + m_{D^\ast}^2] -
  \partial^\mu\partial^\nu ) D^\ast_\nu(x)   \nonumber\\[1mm]
&+&  D^{\ast \, +}_{s \, \mu} (x) ( g^{\mu\nu}
[\Box + m_{D_s^\ast}^2] - \partial^\mu\partial^\nu )
D^{\ast \, -}_{s \, \nu}(x) \, \label{free1-} \,.
\en
Here $\Box = \partial^\mu \partial_\mu$,
$\vec\pi$ is the triplet of pions, $D_s^\pm$ and $D_s^{\ast \, \pm}$
are the pseudoscalar and vector charm-strange mesons, respectively.
The doublets of vector mesons $D^\ast$ and $K^\ast$ are given by
\eq
D^\ast =
\left(
\begin{array}{c}
D^{\ast \, 0} \\
D^{\ast \, +} \\
\end{array}
\right)\,, \hspace*{1cm}
K^\ast =
\left(
\begin{array}{c}
K^{\ast \, +} \\
K^{\ast \, 0} \\
\end{array}
\right)\,.
\en
In our convention the isospin-symmetric meson masses of the iso-multiplets
are identified with the masses of the charged partners~\cite{Yao:2006px}:
\eq
& &m_\pi \equiv m_{\pi^\pm} = 139.57018 \ {\rm MeV}\,, \hspace*{.3cm}
m_K \equiv m_{K^\pm} = 493.677 \ {\rm MeV}\,,          \hspace*{.3cm}
m_{K^\ast} \equiv m_{K^{\ast \, \pm}} = 891.66 \ {\rm MeV}\,, \\
& &m_D \equiv m_{D^\pm} = 1.8693 \ {\rm GeV}\,,        \hspace*{.75cm}
m_{D^\ast} \equiv m_{D^{\ast \, \pm}} = 2.010 \ {\rm GeV}\,. \nonumber
\en
The masses of the iso-singlet states are~\cite{Yao:2006px}:
\eq
& &m_\eta   = 547.51 \ {\rm MeV}\,, \hspace*{1.7cm}
m_{D_s}  =  m_{D^\pm_s} = 1.9682 \ {\rm GeV}\,, \\
& &m_{D^\ast_s} = m_{D^{\ast \, \pm}_s} = 2.112 \ {\rm GeV}\,,
\hspace*{.3cm}
m_{D^\ast_{s0}} = m_{D^{\ast \, \pm}_{s0}} = 2.3173 \ {\rm GeV}\,.
\nonumber
\en
The interaction term ${\cal L}_{\rm int}(x)$ will be discussed later.
First we would like to write down the isospin-breaking term
${\cal L}_{\rm break}$, which includes the mass corrections of the
neutral mesons containing $u$ or $d$ quarks and the $\eta-\pi^0$
mass mixing~\cite{Gross:1979ur,Cho:1994zu}: 
\eq
{\cal L}_{\rm break}(x) \, = \,
\delta {\cal L}^{P}(x) + \delta {\cal L}^{V}(x) + {\cal L}_{\eta\pi}(x) \,,
\en
where
\eq
\delta {\cal L}^{P}(x) &=& \frac{\delta_\pi}{2} \, [\pi^0(x)]^2
+ \delta_K \, \bar K^0(x) K^0(x) + \delta_D \, \bar D^0(x) D^0(x)
\,, \label{delta_P}\\[2mm]
\delta {\cal L}^{V}(x) &=&
- \delta_{K^\ast} \, \bar  K^{\ast \, 0}_\mu (x) K^{\ast \, 0 \, \mu} (x)
- \delta_{D^\ast} \, \bar  D^{\ast \, 0}_\mu (x) D^{\ast \, 0 \, \mu} (x)
\,, \label{delta_V}\\[2mm]
{\cal L}_{\eta\pi}(x) &=& B \,
\frac{m_d - m_u}{\sqrt{3}} \, \pi^0(x) \, \eta(x) \,,
\en
where $m_u$ and $m_d$ are the $u$ and $d$ current quark masses,
$B$ is the condensate parameter. Here $\delta_M$ are the isospin-breaking
parameters which are fixed by the difference of masses squared of the
charged and neutral members of the iso-multiplets as:
\eq
\delta_M  = m_{M^\pm}^2 - m_{M^0}^2\,, \hspace*{.5cm}
 m_{M^0} \equiv m_{\bar M^0}\,.
\en
The set of $m_{M^0}$ is taken from data~\cite{Yao:2006px} with:
\eq
& &m_{\pi^0} = 134.9766 \ {\rm MeV}\,, \hspace*{.3cm}
m_{K^0} = 497.648 \ {\rm MeV}\,,    \hspace*{.3cm}
m_{K^{\ast \, 0}} = 896.0 \ {\rm MeV}\,, \\
& &m_{D^0} = 1.8645 \ {\rm GeV}\,, \hspace*{.3cm} \hspace*{.3cm}
m_{D^{\ast \, 0}} = 2.0067 \ {\rm GeV}\,. \nonumber
\en
Eqs.~(\ref{free0+})-(\ref{free1-}), (\ref{delta_P}) and (\ref{delta_V})
define the free meson propagators for scalar (pseudoscalar) fields
\eq
i \, D_M(x-y) = \langle 0 | T \, M(x) \, M^\dagger(y)  | 0 \rangle
\ = \
\int\frac{d^4k}{(2\pi)^4i} \, e^{-ik(x-y)} \ \tilde D_M(k) \,,
\en
where
\eq
\tilde D_M(k) = \frac{1}{m_M^2 - k^2 - i\epsilon}
\en
and vector fields
\eq
i \, D_{M^\ast}^{\mu\nu}(x-y) = \langle 0 | T \, M^{\ast \, \mu}(x) \,
M^{\ast \, \nu \, \dagger}(y) | 0 \rangle \ = \
\int\frac{d^4k}{(2\pi)^4i} \, e^{-ik(x-y)} \
\tilde D_{M^\ast}^{\mu\nu}(k)
\en
where
\eq
\tilde D_{M^\ast}^{\mu\nu}(k) = - \frac{1}{m_{M^\ast}^2 - k^2 - i\epsilon}
\biggl( g^{\mu\nu} - \frac{k^\mu k^\nu}{m_{M^\ast}^2} \biggr) \,.
\en
In the following calculations it will be convenient to expand the
propagators of the neutral mesons $D^0 (\bar D^0)$, $K^0 (\bar K^0)$,
$D^{\ast \, 0} (\bar D^{\ast \, 0})$ and
$K^{\ast \, 0} (\bar K^{\ast \, 0})$ in powers of the
corresponding isospin-breaking parameters as:
\eq
\tilde D_{M^0}(k) &=& \biggl[ 1 - \delta_M
\frac{\partial}{\partial m_{M^{\pm}}^2} \biggr] \tilde D_{M^\pm}(k)
+ O(\delta_M^2) \,, \\
\tilde D_{M^{\ast \, 0}}^{\mu\nu}(k) &=& \biggl[ 1 - \delta_{M^\ast}
\frac{\partial}{\partial m_{M^{\ast \pm}}^2} \biggr]
\tilde D_{M^{\ast \, \pm}}^{\mu\nu}(k) + O(\delta_{M^\ast}^2) \,. \nonumber
\en
The interaction Lagrangian includes the strong and electromagnetic parts
\eq\label{L_int}
  {\cal L}_{\rm int}(x) = {\cal L}_{\rm int}^{\rm str}(x)
+ {\cal L}_{\rm int}^{\rm em}(x) \,,
\en
as already apparent from the previous discussion related to Figs.2-4.  
The relevant strong part of the effective Lagrangian contains the
following terms: the Lagrangian ${\cal L}_{D_{s0}^\ast}$~(\ref{Lagr_Ds0})
describing  the coupling of the $D_{s0}^\ast$ meson to its constituents
and $VPP$-type Langrangians, describing the interaction of vector mesons
with two pseudoscalars: 
\eq\label{L_int_str}
{\cal L}_{\rm int}^{\rm str}(x) &=&
{\cal L}_{D_{s0}^\ast}(x)
+ {\cal L}_{D^\ast D \pi}(x) + {\cal L}_{D^\ast D \eta}(x)
+ {\cal L}_{K^\ast K \pi}(x) + {\cal L}_{K^\ast K \eta}(x)  \nonumber\\
&+& {\cal L}_{D^\ast D_s K}(x) + {\cal L}_{K^\ast D_s D}(x)
+ {\cal L}_{D^\ast_s D K}(x) \,.
\en
Let us specify the $VPP$ interaction Lagrangians occurring in
Eqs.~(\ref{L_int_str}). In general they can be defined as:
\eq\label{L_str_VPP} 
{\cal L}_{VP_1P_2}(x) \, = \, g_{_{VP_1P_2}} \,
V_\mu(x) \, P_1(x) \,
i\!\stackrel{\leftrightarrow}{\partial}^{\,\mu}
P_2(x)  \, + \, {\rm H.c.} 
\en 
To be consistent with the definitions occurring in literature, we
use the following form of the particular Lagrangians:
\eq 
{\cal L}_{D^\ast D \pi}(x) &=&
- \frac{g_{_{D^\ast D \pi}}}{2\sqrt{2}} \,
D^{\ast \, \dagger}_\mu(x) \, \vec\tau \, \vec{\pi}(x)
i\!\stackrel{\leftrightarrow}{\partial}^{\,\mu} \! D(x)
\, + \, {\rm H.c.}
\, \\
{\cal L}_{D^\ast D \eta}(x) &=&
- \frac{g_{_{D^\ast D \eta}}}{2\sqrt{2}} \,
D^{\ast \, \dagger}_\mu(x) \, \eta(x) \,
i\!\stackrel{\leftrightarrow}{\partial}^{\,\mu} \! D(x)
\, + \, {\rm H.c.}  \, \\
{\cal L}_{K^\ast K \pi}(x) &=& \frac{g_{_{K^\ast K \pi}}}{\sqrt{2}} \,
K^{\ast \, \dagger}_\mu(x) \, \vec\tau \, \vec{\pi}(x)
i\!\stackrel{\leftrightarrow}{\partial}^{\,\mu} \! K(x)  \, + \, {\rm H.c.}
\, \\
{\cal L}_{K^\ast K \eta}(x) &=& \frac{g_{_{K^\ast K \eta}}}{\sqrt{2}} \,
K^{\ast \, \dagger}_\mu(x) \, \eta(x) \,
i\!\stackrel{\leftrightarrow}{\partial}^{\,\mu} \! K(x)  \, + \, {\rm H.c.}
\, \\
{\cal L}_{D^\ast D_s K}(x) &=& g_{_{D^\ast D_s K}} \,
D^{\ast \, T}_\mu(x) \, K(x)
\, i\!\stackrel{\leftrightarrow}{\partial}^{\,\mu} \! D_s^-(x)
\, + \, {\rm H.c.}  \, \\
{\cal L}_{K^\ast D_s D}(x) &=&  g_{_{K^\ast D_s D}} \,
K^{\ast \, T}_\mu(x) D(x)
\, i\!\stackrel{\leftrightarrow}{\partial}^{\,\mu} \! D_s^-(x)
\,
\, + \, {\rm H.c.}  \, \\
{\cal L}_{D^\ast_s D K}(x) &=& g_{_{D^\ast_s D K}} \,
D^{\ast \, -}_{s \, \mu}(x) \, D^T(x)
\, i\!\stackrel{\leftrightarrow}{\partial}^{\,\mu} \! K(x)
\, + \, {\rm H.c.}  \, 
\en 
where summation over isospin indices is understood 
and $A \stackrel{\leftrightarrow}{\partial} B
\equiv A \partial B - B \partial A$. 

The couplings $g_{_{D^\ast D \pi}}$ and $g_{_{K^\ast K \pi}}$
are fixed by data for the strong decay widths
$D^\ast \to D \pi$ and $K^\ast \to K \pi$.
In particular, the strong two-body decay widths
$\Gamma(D^{\ast \, +} \to  D^0 \pi^+)$ and
$\Gamma(K^{\ast \, +} \to  K^0 \pi^+)$ are related to
$g_{_{D^\ast D \pi}}$~\cite{Belyaev:1994zk,Anastassov:2001cw}
and $g_{_{K^\ast K \pi}}$
as
\eq
\Gamma(D^{\ast \, +} \to  D^0 \pi^+) &=&
\frac{g_{_{D^\ast D \pi}}^2}{24 \pi m_{D^{\star +}}^2} \,
P_{_{\pi D^{\ast}}}^3 \,, \\
\Gamma(K^{\ast \, +} \to  K^0 \pi^+) &=&
\frac{g_{_{K^\ast K \pi}}^2}{6 \pi m_{K^{\star +}}^2} \,
P_{_{\pi K^{\ast}}}^3 \,,
\en
where $P_{_{\pi V}}$ is the three-momentum of $\pi^+$ in the
rest frame of the decaying vector meson $V$. Using data for the
corresponding strong decay widths one deduces: 
$g_{_{D^\ast D \pi}} = 17.9$~\cite{Anastassov:2001cw} 
and $g_{_{K^\ast K \pi}} = 4.61$~\cite{Yao:2006px}. 

The coupling constants $g_{_{D^\ast D \pi(\eta)}}$  
are obtained in the context of heavy hadron chiral perturbation
theory (HHChPT)~\cite{Wise:1992hn}. The couplings
$g_{_{D^\ast D \pi}}$ and $g_{_{D^\ast D \eta}}$ are expressed
(and then related) in terms of a universal strong coupling constant
$g$ involving heavy (vector and pseudoscalar) and Goldstone
mesons and in terms of the leptonic decay constants $F_P$:
\eq\label{DsDpi_DsDeta}
g_{_{D^\ast D \pi}} = \frac{m_{D^\ast}}{F_\pi} g \,  \sqrt{2}\,,
\hspace*{.3cm}
g_{_{D^\ast D \eta}} = \frac{m_{D^\ast}}{F_\eta} g  \,
\sqrt{\frac{2}{3}}\,,
\en
where $F_\pi = 92.4$ MeV and $F_\eta = 1.3 \, F_\pi$.
From Eq.~(\ref{DsDpi_DsDeta}) and using $g_{_{D^\ast D \pi}} = 17.9$ 
we deduce the value of $g_{_{D^\ast D \eta}}$ with
\eq
 g_{_{D^\ast D \eta}} \, = \, \frac{F_\pi}{F_\eta \sqrt{3}}
 g_{_{D^\ast D \pi}}  \, = \, 7.95 \,.
\en
The coupling constant $g_{_{K^\ast K \eta}}$ can be related to
$g_{_{K^\ast K \pi}}$ using the unitary symmetry relation:
\eq
g_{_{K^\ast K \eta}} \, = \, \frac{F_\pi \sqrt{3}}{F_\eta}
 g_{_{K^\ast K \pi}}  \, = \, 6.14 \,.
\en
Again, as in the case of $g_{_{D^\ast D \pi(\eta)}}$, we include in
couplings the relation to the corresponding decay constants
$F_\pi$ and $F_\eta$.
 
The coupling constants $g_{_{D^\ast D_s K}}$ and $g_{_{D_s^\ast D K}}$
have been estimated using the QCD sum rule technique
in Refs.~\cite{Wang:2006id,Bracco:2006xf}. These couplings are
important for the evaluation of the dissociation cross section
of $J/\Psi$ to kaons (see, e.g. discussion in
Refs.~\cite{Haglin:1999xs,Azevedo:2003qh}.). Here we use the results
of Ref.~\cite{Wang:2006id}:
$g_{_{D^\ast D_s K}} = 2.02$ and  $g_{_{D_s^\ast D K}} = 1.84$.
The coupling $g_{_{K^\ast D_s D}}$ can also be related
to $g_{_{D^\ast D_s K}}$, using $SU(4)$ symmetry arguments:
$g_{_{K^\ast D_s D}} = g_{_{D^\ast D_s K}} = 2.02$. 

The relevant electromagnetic part has three main terms:
\eq\label{L_int_em}
{\cal L}_{\rm int}^{\rm em}(x)
= {\cal L}_{\rm int}^{\rm em(1)}(x)
+ {\cal L}_{\rm int}^{\rm em(2)}(x)
+ {\cal L}_{\rm int}^{\rm em(3)}(x) \,. 
\en
The first term describes the local coupling of charged $D$-, $K$-
and $D_s^\ast$ mesons to the electromagnetic field
\eq\label{L_em_1}
{\cal L}^{\rm em (1)}_{\rm int}(x) &=& i e A_\mu(x) \, \biggl\{  \,
D^-(x) \stackrel{\leftrightarrow}{\partial}^{\,\mu} \! D^+(x)
+ K^-(x) \stackrel{\leftrightarrow}{\partial}^{\,\mu} \!  K^+(x) \,
\nonumber\\
&-& D_s^{\ast - \alpha}(x) \stackrel{\leftrightarrow}{\partial}^{\,\mu} \!
D_{s \alpha}^{\ast +}(x) + \frac{1}{2}
D_s^{\ast - \alpha}(x) \stackrel{\leftrightarrow}{\partial}_{\,\alpha} \!
D_s^{\ast + \mu}(x)
+ \frac{1}{2}
D_s^{\ast - \mu}(x) \stackrel{\leftrightarrow}{\partial}^{\,\alpha} \!
D_{s \alpha}^{\ast +}(x) \biggr\} \,. 
\en 
The term ${\cal L}^{\rm em (1)}_{\rm int}$ is generated after gauging 
of the free meson Lagrangians using minimal substitution:
\eq
\partial^\mu M^\pm \, \to \, (\partial^\mu \mp ie A^\mu) \, M^\pm \,.
\en
The terms ${\cal L}_{\rm int}^{\rm em(2)}$ and
${\cal L}_{\rm int}^{\rm em(3)}$ are generated due to the gauging of
the strong Lagrangians (\ref{L_str_VPP}) and (\ref{Lagr_Ds0}) containing
derivatives acting on the charged fields. Note, that the correlation
function $\Phi_{D_{s0}^\ast}$, describing the nonlocal $D_{s0}^\ast DK$
coupling, is a function of $\partial^2$ and, therefore,
both Lagrangians (\ref{L_str_VPP}) and (\ref{Lagr_Ds0})
are not gauge-invariant under electromagnetic $U_{\rm em}(1)$
transformations and should be modified accordingly.

To get the second term we replace all derivatives acting on the charged
fields by the covariant ones using minimal substitution
(as is the case for gauging the free Lagrangians). The term in
${\cal L}^{\rm em (2)}$ relevant for our calculation in contains
the coupling of the vector $D_s^\ast$ meson to $D$, $K$ and the 
photon field with
\eq\label{L_em_2}
{\cal L}^{\rm em (2)}_{\rm int}(x) &=&  e \, g_{_{D^\ast_s D K}} \,
A^\mu(x) \, D^{\ast \, -}_{s \, \mu}(x) \, [ \, D^0(x) \, K^+(x) \, - \,
D^+(x) \, K^0(x) \, ] \, + \, {\rm H.c.}  \, + \, \cdots
\en
The gauging of the nonlocal Lagrangian of Eq.~(\ref{Lagr_Ds0})
proceeds in a way suggested in Ref.~\cite{Mandelstam:1962mi}
and extensively used in Refs.~\cite{Efimov:1987na,Faessler:2003yf}.
In particular, to guarantee local invariance of the strong interaction
Lagrangian, in ${\cal L}_{\rm int}^{\rm str}$ each charged constituent
meson field (i.e. $D^\pm$ and $K^\pm$ meson fields) are multiplied by
the gauge field exponential resulting in
\eq\label{L_str_gauging}
{\cal L}_{\rm int}^{\rm str + em(3)}(x) &=& g_{_{D_{s0}^\ast}} \,
D_{s0}^{\ast \, -}(x) \, \int\! dy \, \Phi_{D_{s0}^\ast}(y^2) \,
\biggl\{ e^{- i e I(x+w_Ky,x,P)} D^+(x+w_K y) K^0(x-w_D y) \, \nonumber\\
&+& D^0(x+w_K y) e^{- i e I(x-w_Dy,x,P)} K^+(x-w_D  y) \biggr\}
\, + \, {\rm H.c.} 
\en
where
\eq\label{path}
I(x,y,P) = \int\limits_y^{x} dz_\mu A^\mu(z).
\en
For the derivative of the path integral (\ref{path}) we use the
path-independent prescription suggested in Refs.~\cite{Mandelstam:1962mi}
\eq\label{path1}
\lim\limits_{dx^\mu \to 0} dx^\mu
\frac{\partial}{\partial x^\mu} I(x,y,P) \, = \,
\lim\limits_{dx^\mu \to 0} [ I(x + dx,y,P^\prime) - I(x,y,P) ] \,,
\en
where path $P^\prime$ is obtained from $P$ when shifting the end-point $x$
by $dx$. Use of the definition~(\ref{path1})
leads to the key rule
\begin{eqnarray}\label{path2}
\frac{\partial}{\partial x^\mu} I(x,y,P) = A_\mu(x) \,,
\end{eqnarray}
which in turn states that the derivative of the path integral $I(x,y,P)$ does
not depend on the path P originally used in the definition. The non-minimal
substitution (\ref{L_str_gauging}) is therefore completely equivalent to the
minimal prescription.

In the calculation of the amplitudes of the radiative
$D_{s0}^{\ast} \to D_s^\ast \gamma$ decay,
in Eq.~(\ref{L_str_gauging}) we only need to keep terms linear in
$A_\mu$, that is the four-particle coupling
$D_{s0}^\ast DK\gamma$. Hence, the third term contributing to the
electromagnetic interaction Lagrangian is given by
\eq\label{L_em_3}
{\cal L}_{\rm int}^{\rm em(3)}(x) &=& - \, i e g_{_{D_{s0}^\ast}} \,
D_{s0}^{\ast \, -}(x) \, \int\! dy \, \Phi_{D_{s0}^\ast}(y^2) \,
\biggl\{ \int\limits_x^{x+w_Ky} dz_\mu A^\mu(z)
D^+(x+w_K y) K^0(x-w_D y) \,\\
&+& \int\limits_x^{x-w_Dy} dz_\mu A^\mu(z)
D^0(x+w_K y) K^+(x-w_D  y) \biggr\}
\, + \, {\rm H.c.} + \cdots \nonumber
\en
Concluding the discussion of the effective interaction Lagrangian  
we stress that all couplings occurring in the diagrams contributing  
to the decays $D_{s0}^{\ast} \to D_s \pi^0$ and  
$D_{s0}^{\ast} \to D_s^{\ast} \gamma$ are explicitly fixed,  
except $g_{_{D_{s0}^\ast}}$ discussed in the following. 

\subsection{Analysis of the $D_{s0}^\ast DK$ coupling $g_{_{D_{s0}^\ast}}$}

Finally, we discuss the numerical value of the model-dependent
constant $g_{_{D_{s0}^\ast}}$. In terms of a general functional
form of the correlation function ${\widetilde{\Phi}}_{D_{s0}^\ast}$
the coupling constant $g_{_{D_{s0}^\ast}}$ is given by:
\eq
\frac{1}{g_{_{D_{s0}^\ast}}^2} =
\frac{2}{(4 \pi \Lambda_{D_{s0}^\ast})^2} \,
\int\limits_0^\infty \int\limits_0^\infty
\frac{R \, d\alpha_1 d\alpha_2}{(1 + \alpha_1 + \alpha_2)^3}
\,\, [ - d\tilde \Phi^2_{D_{s0}^\ast}(z)/dz] \,,
\en
where
\eq
z &=& \mu_D^2 \alpha_1 + \mu_K^2 \alpha_2 
   - \frac{R \, \mu_{D_{s0}^\ast}^2}{1 + \alpha_1 + \alpha_2} \,, \\
R &=& \alpha_1 \alpha_2 + \alpha_1 w_D^2 + \alpha_2 w_K^2\,, 
\hspace*{.5cm} \mu_M = \frac{m_M}{\Lambda_{D_{s0}^\ast}} \,.
\nonumber
\en
One should stress that coupling constant $g_{_{D_{s0}^\ast}}$ remains
finite when we remove the cutoff $\Lambda_{D_{s0}^\ast} \to \infty$
(local limit). A finite result is obtained, because the derivative of
the $D_{s0}^\ast$ mass operator is convergent, i.e. the loop integral is
$\int d^4k/k^6$ when the correlation function
$\tilde \Phi_{D_{s0}^\ast}$ is removed (or equal to one) at
$\Lambda_{D_{s0}^\ast} \to \infty$.
However, in the calculation of transition diagrams
(like in Figs.2-4) we deal with divergent integral and, therefore,
we need the correlation function to perform the regularization of the
occurring loop integrals. Now the question is how sensitive our results
are to a variation of $\Lambda_{D_{s0}^\ast}$. First, we look at the
coupling constant $g_{_{D_{s0}^\ast}}$. In the limit
$\Lambda_{D_{s0}^\ast} \to \infty$  it is given by
\eq
\frac{1}{g_{_{D_{s0}^\ast}}^2} = \frac{2}{(4 \pi
m_{D_{s0}^\ast})^2} \,
\biggl\{ \frac{m_D^2 - m_K^2}{m_{D_{s0}^\ast}^2} \,
{\rm ln}\frac{m_D}{m_K} \, - 1 \, + \,
\frac{m_{D_{s0}^\ast}^2 (m_D^2+m_K^2)
- (m_D^2 - m_K^2)^2}{m_{D_{s0}^\ast}^2\sqrt{-\lambda}} \,
\sum\limits_\pm {\rm arctan}\frac{z_\pm}{\sqrt{-\lambda}} \biggr\}
\en
where $z_\pm = m_{D_{s0}^\ast}^2 \pm (m_D^2 - m_K^2)$ and
\eq
\lambda \doteq \lambda(m_{D_{s0}^\ast}^2,m_D^2,m_K^2) =
m_{D_{s0}^\ast}^4 + m_D^4 + m_K^4
- 2 m_{D_{s0}^\ast}^2 m_D^2
- 2 m_{D_{s0}^\ast}^2 m_K^2
- 2 m_D^2 m_K^2
\en
is the K\"allen function.

For checking purposes we also analyze the coupling
$g_{_{D_{s0}^\ast}}$ in the heavy quark limit (HQL),
where the masses of $D$ and $D_{s0}^\ast$ mesons together
with the charm quark mass $m_c$ go to infinity. In the HQL
the $D$ meson in the $D_{s0}^\ast$ molecule fixes
the center-of-mass, surrounded by a light $K$ meson
in analogy to the heavy-light $Q \bar q$ mesons.
For the nonlocal case the result for $g_{_{D_{s0}^\ast}}$
in the HQL is:
\eq\label{HQL_nonl}
\frac{1}{g_{_{D_{s0}^\ast}}^2} &=&
\frac{1}{(4 \pi m_c)^2} \, \int\limits_0^\infty
\frac{\displaystyle{d\alpha}}
{\displaystyle{1 + \mu_K^2 \alpha}} \,\,
\Phi^2_{D_{s0}^\ast}(\alpha) \,.
\en
The HQL result for the local case is:
\eq\label{HQL_loc}
\frac{1}{g_{_{D_{s0}^\ast}}^2} =
\frac{1}{(4 \pi m_c)^2} \, {\rm ln}\frac{m_c^2}{m_K^2} \,.
\en
Now we compare our numerical results for the coupling constant
$g_{_{D_{s0}^\ast}}$ in different regimes: 1)~nonlocal case (NC); 
2)~local case (LC); 3)~nonlocal case + HQL (NCHQL) and
4)~local case + HQL (LCHQL). When we deal with the nonlocal case
we proceed with the Gaussian correlation function
$\Phi_{D_{s0}^\ast}(z) = {\rm exp}(-z)$ and vary the
scale parameter $\Lambda_{D_{s0}^\ast}$  from 1 to 2 GeV.
For the charm quark mass we use the averaged result of the
PDG~\cite{Yao:2006px}: $m_c = 1.25$ GeV.
For a convenience we attach a corresponding superscript to
$g_{_{D_{s0}^\ast}}$ to indicate the specific regime.

We get the following results: the
coupling $g_{_{D_{s0}^\ast}}^{\rm NC}$  varies from 11.26 GeV
at $\Lambda_{D_{s0}^\ast} = 1$ GeV to 9.90 GeV
at $\Lambda_{D_{s0}^\ast} = 2$~GeV.
The coupling $g_{_{D_{s0}^\ast}}^{\rm LC}$ is expressed only
in terms of physical meson masses with the result
$g_{_{D_{s0}^\ast}}^{\rm LC} = 8.98$ GeV.
The coupling $g_{_{D_{s0}^\ast}}^{\rm NCHQL}$  varies from 16.22 GeV
at $\Lambda_{D_{s0}^\ast} = 1$ GeV to 11.52 GeV
at $\Lambda_{D_{s0}^\ast} = 2$ GeV.
Finally, we have $g_{_{D_{s0}^\ast}}^{\rm LCHQL} = 11.52$ GeV.
All results for $g_{_{D_{s0}^\ast}}$ are quite close to each other
with a typical value for $g_{_{D_{s0}^\ast}}$ of about 10 GeV which
is consistent with preceding calculations done in other theoretical
approaches. In Table 1 we compare our result for the $D_{s0}^\ast DK$
coupling constant to predictions of other theoretical approaches
(we use a compilation of the results done in Ref.~\cite{Wang:2006id}).

\subsection{Extension to other possible hadronic molecules}

We end this section with a comment concerning the extension
of the derived framework to the study of other hadronic molecules.
This can be done in a straightforward fashion. The starting point
is the construction of an effective Lagrangian describing hadronic
molecules as bound states of its constituents. In particular, for
the case of the charm-strange meson $D_{s1}(2460)$ and for the
possible partners in the bottom sector $B_{s0}^\ast(5725)$ and
$B_{s1}(5778)$ the simplest Lagrangians have the form: 
\eq
{\cal L}_{D_{s1}}(x) &=& g_{_{D_{s1}}} \,
D_{s1}^{- \, \mu}(x) \, \int\! dy \,
\Phi_{D_{s1}}(y^2) \, D^{\ast \, T}_\mu(x+w_{_{KD^\ast}} y) \,
K(x-w_{_{D^\ast K}} y) \, + \, {\rm H.c.} \,, \label{Lagr_Ds1} \\
{\cal L}_{B_{s0}^\ast}(x) &=& g_{_{B_{s0}^\ast}} \,
\bar B_{s0}^{\ast \, 0}(x) \, \int\! dy \,
\Phi_{B_{s0}^\ast}(y^2) \, B^\dagger(x+w_{_{KB}} y) \, K(x-w_{_{BK}} y)
\, + \, {\rm H.c.} \,, \label{Lagr_Bs0} \\
{\cal L}_{B_{s1}}(x) &=& g_{_{B_{s1}}} \,
\bar B_{s1}^{0 \, \mu}(x) \, \int\! dy \,
\Phi_{B_{s1}}(y^2) \, B^{\ast \, \dagger}_\mu(x+w_{_{KB^\ast}} y) \,
K(x-w_{_{B^\ast K}} y)
\, + \, {\rm H.c. } \,, \label{Lagr_Bs1}
\en
where $w_{ij} = m_i/(m_i + m_j)$, $g_M$ and $\Phi_M$ are the
coupling constants (fixed from the compositeness condition)
and correlation functions. The doublets of $B^{(\ast)}$ and 
$B^{(\ast) \, \dagger}$ mesons are defined as
\eq
B^{(\ast)} =
\left(
\begin{array}{c}
B^{(\ast) \, +} \\
B^{(\ast) \, 0} \\
\end{array}
\right)\,, \hspace*{1cm}
B^{(\ast) \, \dagger} = 
(B^{(\ast) \, -} \  \ \bar B^{(\ast) \, 0} ) \,. 
\en
The molecular structure of $D_{s1}^\pm$, 
$B_{s0}^{\ast \, 0}$, $\bar B_{s0}^{\ast \,0}$,
$B_{s1}^0$ and $\bar B_{s1}^0$ is:
\eq
& &|D_{s1}^+\ra \, = \, |D^{\ast \, +} K^0\ra
                      + |D^{\ast \, 0} K^+\ra \,, \hspace*{1cm}
   |D_{s1}^-\ra \, = \, |D^{\ast \, -} \bar K^0\ra
                      + |\bar D^{\ast \, 0} K^-\ra \,,
\nonumber\\
& &|B_{s0}^{\ast 0}\ra \, = \, |B^+ K^-\ra
                      + |B^0 \bar K^0\ra \,, \hspace*{1.4cm}
   |\bar B_{s0}^{\ast 0}\ra \, = \, |B^- K^+\ra
                        +    |\bar B^0 K^0 \ra \, \\
& &|B_{s1}^0\ra \, = \, |B^{\ast \, +} K^-\ra
                      + |B^{\ast \, 0} \bar K^0\ra \,, \hspace*{1.05cm}
   |\bar B_{s1}^0\ra \, = \, |B^{\ast \, -} K^+\ra
                      + |\bar B^{\ast 0} K^0\ra \,. \nonumber
\en
The calculation of decay properties of $D_{s1}(2460)$, $B_{s0}^\ast(5725)$
and $B_{s1}(5778)$ mesons goes beyond the scope of the present paper
and we relegate this issue to a forthcoming paper.

\section{Strong $D_{s0}^{\ast} \to D_s \pi^0$ and
radiative $D_{s0}^{\ast} \to D_s^{\ast} \gamma$ decays}

In this section we discuss the numerical results
for the $D_{s0}^{\ast} \to D_s \pi^0$ and
$D_{s0}^{\ast} \to D_s^{\ast} \gamma$ decay properties.
As we already stressed in the preceding section two types of
diagrams contribute to the amplitude of the strong decay
$D_{s0}^{\ast} \to D_s \pi^0$: the ``direct'' diagrams of Fig.2
and the ``$\eta-\pi^0$ mixing'' diagrams of Fig.3. The ``direct''
diagrams occur due to the $DK$ molecular structure of the
$D_{s0}^{\ast}$ meson, while in the two-quark picture they are
forbidden according to the Okubo-Zweig-Iizuka rule.
In the framework of our approach this is not the case,
since $D$ and $K$ mesons contain nonstrange quarks.
The total contribution of the ``direct'' diagrams starts
at order $O(\delta)$, where $\delta$ of Eq.~(\ref{delta_par}) is
the generic parameter of isospin breaking.
Hence, the ``direct diagrams'' are of the same order
in the isospin-breaking counting scheme as the ``$\eta-\pi^0$ mixing''
diagrams, and, therefore, both types of diagrams should be included.
To clarify this mechanism we present our results our results for
two cases: 1) ``full calculation'' (Full) and
2) ``leading-order'' (LO), i.e.
restricting to first order in isospin-breaking $O(\delta)$.

It is convenient to write the matrix element describing
the $D_{s0}^{\ast} \to D_s \pi^0$ transition as a sum of
the added contributions of the diagrams in Figs.2 and 3:
\eq
M(D_{s0}^{\ast} \to D_s \pi^0) \, = \,
M_{\rm dir}(D_{s0}^{\ast} \to D_s \pi^0)
\, + \, M_{\rm mix}(D_{s0}^{\ast} \to D_s \pi^0)
\en
with
\eq
& &M_{\rm dir}(D_{s0}^{\ast} \to D_s \pi^0)  =
g_{D_{s0}^{\ast} D_s \pi^0} \,, \label{MDs0Dspi}\\
& &M_{\rm mix}(D_{s0}^{\ast} \to D_s \pi^0)  =
g_{D_{s0}^{\ast} D_s \eta} \frac{m_d - m_u}{m_s - \hat m}
\frac{\sqrt{3}}{4}
\,, \label{MDs0Dseta}
\en
where $\hat m = (m_u + m_d)/2$ and
$(m_d - m_u)/(m_s - \hat m) = 1/43.7$ (see e.g. Ref.~\cite{Cho:1994zu}).
In the derivation of Eq.~(\ref{MDs0Dseta}) we use the masses of
$\pi^0$ and $\eta$ meson in leading order of the chiral expansion.
The total effective $D_{s0}^{\ast} D_s \pi^0$ coupling, denoted as
$G_{D_{s0}^{\ast} D_s \pi}$ includes both contributions of the set of
diagrams of Figs.2 and 3 with
\eq
& &G_{D_{s0}^{\ast} D_s \pi}  \, = \,
G_{D_{s0}^{\ast} D_s \pi}^{\rm dir}
+ G_{D_{s0}^{\ast} D_s \pi}^{\rm mix}  \,, \\
& &G_{D_{s0}^{\ast} D_s \pi}^{\rm dir} \equiv
g_{D_{s0}^{\ast} D_s \pi}\,, \hspace*{.5cm}
G_{D_{s0}^{\ast} D_s \pi}^{\rm mix} \equiv
g_{D_{s0}^{\ast} D_s \eta} \frac{m_d - m_u}{m_s - \hat m}
\frac{\sqrt{3}}{4} \,.
\en
In terms of $G_{D_{s0}^\ast D_s \eta}$ the $D_{s0}^{\ast} \to D_s \pi^0$
decay width reads as:
\eq
\Gamma(D_{s0}^{\ast} \to D_s \pi) \, = \,
\frac{G_{D_{s0}^{\ast} D_s \pi}^2}{8\pi m_{D_{s0}^\ast}^2} \, 
P^\ast_{\pi^0} \,, 
\en
where
$P^\ast_{\pi^0} =
\lambda^{1/2}(m_{D_{s0}^\ast}^2,m_{D_s}^2,m_{\pi^0}^2)/(2 m_{D_{s0}^\ast})$ 
is the three-momentum of the decay products. 

The matrix element describing the $D_{s0}^{\ast} \to D_s^{\ast} \gamma$
transition can be written in the manifestly gauge-invariant form
\eq
M_{\mu\nu}(D_{s0}^{\ast} \to D_s^{\ast} \gamma) \, = \,
e \, G_{D_{s0}^{\ast} D_s^{\ast} \gamma} \,
( g_{\mu\nu} p^\prime q - p^\prime_\mu q_\nu ) \,,
\en
where $p^\prime$ and $q$ are the $D_s^\ast$ and photon four-momenta and
$p = p^\prime + q$ is the $D_{s0}^{\ast}$ momentum. Here
$G_{D_{s0}^{\ast} D_s^{\ast} \gamma}$ is the effective
$D_{s0}^{\ast} D_s^{\ast} \gamma$ coupling constant and the
$D_{s0}^{\ast} \to D_s^{\ast} \gamma$ decay width is given by
\eq
\Gamma(D_{s0}^{\ast} \to D_s^{\ast} \gamma) \, = \, \alpha \,
G_{D_{s0}^{\ast} D_s^{\ast} \gamma}^2 \, P_\gamma^{\ast \, 3}
\en
where
\eq
P^\ast_{\gamma} \, = \, \frac{m_{D_{s0}^\ast}}{2} \,
\biggl[ 1 - \frac{m_{D_{s}^\ast}^2}{m_{D_{s0}^\ast}^2} \biggr]
\en 
is the three-momentum of the decay products. 

Now we present the numerical results. First we discuss the contributions
of the different diagrams of Figs.2 and 3 to the effective coupling
$G_{D_{s0}^{\ast} D_s \pi}$. With a typical value for  the scale parameter
of $\Lambda_{D_{s0}^\ast} = 1$ GeV we get the following. 

\vspace*{.25cm}

In the Full calculation, 
\eq
& &G_{D_{s0}^{\ast} D_s \pi} = 146.6 \ {\rm MeV} \,,
\hspace*{.3cm}
G_{D_{s0}^{\ast} D_s \pi}^{\rm dir} = 104.5 \ {\rm MeV} \,,
\hspace*{.3cm}
G_{D_{s0}^{\ast} D_s \pi}^{\rm mix} = 42.1 \ {\rm MeV} \,, \\
& &G_{D_{s0}^{\ast} D_s \pi}^{DD^\ast K} = 40.9 \ {\rm MeV} \,,
\hspace*{.3cm}
G_{D_{s0}^{\ast} D_s \pi}^{{\rm dir}, KK^\ast D} = 63.6 \ {\rm MeV} \,,
\hspace*{.3cm}
G_{D_{s0}^{\ast} D_s \pi}^{{\rm mix}, DD^\ast K} = 7.9 \ {\rm MeV} \,,
\hspace*{.3cm}
G_{D_{s0}^{\ast} D_s \pi}^{{\rm mix}, KK^\ast D} = 34.1 \ {\rm MeV} \,. 
\nonumber
\en
\vspace*{.25cm}

and in the LO calculation, 
\eq
& &G_{D_{s0}^{\ast} D_s \pi} = 145.4 \ {\rm MeV} \,,
\hspace*{.3cm}
G_{D_{s0}^{\ast} D_s \pi}^{\rm dir} = 103.4 \ {\rm MeV} \,,
\hspace*{.3cm}
G_{D_{s0}^{\ast} D_s \pi}^{\rm mix} = 42.0 \ {\rm MeV} \,, \\
& &G_{D_{s0}^{\ast} D_s \pi}^{{\rm dir}, DD^\ast K} = 40.2 \ {\rm MeV} \,,
\hspace*{.3cm}
G_{D_{s0}^{\ast} D_s \pi}^{{\rm dir}, KK^\ast D} = 63.2 \ {\rm MeV} \,,
\hspace*{.3cm}
G_{D_{s0}^{\ast} D_s \pi}^{{\rm mix}, DD^\ast K} = 7.9 \ {\rm MeV} \,,
\hspace*{.3cm}
G_{D_{s0}^{\ast} D_s \pi}^{{\rm mix}, KK^\ast D} = 34.1 \ {\rm MeV} \,,
\nonumber
\en
where the superscripts $DD^\ast K$ and  $KK^\ast D$ relate to the
diagrams of Figs.2(a), 2(b), 3(a), and 3(b) and 2(c), 2(d), 3(c), 
and 3(d), respectively. The direct diagrams
dominate over the mixing diagrams by about a factor of 2. The results
for the decay width (total result and partial contributions of the
different diagrams) are as follows. 

\vspace*{.25cm}

In the Full calculation, 
\eq\label{Gamma_Full}
& &\Gamma(D_{s0}^\ast \to D_s \pi) = 46.7 \ {\rm KeV}\,,
\hspace*{.25cm}
\Gamma(D_{s0}^\ast \to D_s \pi)^{\rm dir} = 23.7 \ {\rm KeV} \,,
\hspace*{.25cm}
\Gamma(D_{s0}^\ast \to D_s \pi)^{\rm mix} = 3.8 \ {\rm KeV} \,, \\
& &\Gamma(D_{s0}^\ast \to D_s \pi)^{{\rm dir}, DD^\ast K} =
3.6 \ {\rm KeV} \,,
\hspace*{.25cm}
\Gamma(D_{s0}^\ast \to D_s \pi)^{{\rm dir}, KK^\ast D} =
8.8 \ {\rm KeV} \,, \nonumber\\
& &\Gamma(D_{s0}^\ast \to D_s \pi)^{{\rm mix}, DD^\ast K} =
0.1 \ {\rm KeV} \,,
\hspace*{.25cm}
\Gamma(D_{s0}^\ast \to D_s \pi)^{{\rm mix}, KK^\ast D} =
2.5 \ {\rm KeV} \,. \nonumber
\en
In the LO calculation, 
\eq\label{Gamma_LO}
& &\Gamma(D_{s0}^\ast \to D_s \pi) = 46.6 \ {\rm KeV}\,,
\hspace*{.25cm}
\Gamma(D_{s0}^\ast \to D_s \pi)^{\rm dir} = 23.6 \ {\rm KeV} \,,
\hspace*{.25cm}
\Gamma(D_{s0}^\ast \to D_s \pi)^{\rm mix} = 3.9 \ {\rm KeV} \,, \\
& &\Gamma(D_{s0}^\ast \to D_s \pi)^{{\rm dir}, DD^\ast K} =
3.6 \ {\rm KeV} \,,
\hspace*{.25cm}
\Gamma(D_{s0}^\ast \to D_s \pi)^{{\rm dir}, KK^\ast D} =
8.8 \ {\rm KeV} \,, \nonumber\\
& &\Gamma(D_{s0}^\ast \to D_s \pi)^{{\rm mix}, DD^\ast K} =
0.1 \ {\rm KeV} \,,
\hspace*{.25cm}
\Gamma(D_{s0}^\ast \to D_s \pi)^{{\rm mix}, KK^\ast D} =
2.6 \ {\rm KeV} \,. \nonumber
\en
From Eqs.~(\ref{Gamma_Full}) and (\ref{Gamma_LO}) it is evident
that the restriction to the leading-order in isospin breaking is
very good approximation to the full calculation
(both sets of results practically coincide with
each other). We would like to stress that the
strong decay width $\Gamma(D_{s0}^{\ast} \to D_s \pi)$ is
enhanced in a molecular picture for the $D_{s0}^{\ast}$ meson
as compared to the quarkonium interpretation due to the inclusion of the
direct $\pi^0$ coupling to the $DD^\ast$ or $KK^\ast$ meson pairs.  
This enhancement is particularly present, since the 
``direct'' mode dominates over the ``mixing'' mode. 

On the other hand, when turning to the heavy quark limit the
contribution of the ``direct'' mode becomes much smaller, about $0.4$ KeV,
while the one of the ``mixing'' decreases less to about $1.4$ KeV.
The total result for the decay width is an order of magnitude smaller
as in the full dynamical case with
$\Gamma(D_{s0}^\ast \to D_s \pi) \simeq 3.3$ KeV.
From the results obtained in the HQL we make the following conclusions:
first, in the HQL the ``mixing'' mode dominates over the ``direct'' mode.
This result is consistent with heavy hadron ChPT by
conception (restriction to the ``mixing'' mode) and numerically
(the result for the width is of the order of a few KeV).
Second, we have a clear explanation why in the HQL the ``direct'' mode
is suppressed. The reason is that the isospin breaking effects
due the difference of heavy $D^{(\ast)}$ mesons occurring in the loop
are of next-to-leading order in the $1/m_c$ expansion, i.e. they
are of the form $\delta_{D^{(\ast)}}/(m_c\Lambda_{D_{s0}^\ast})$.
Numerically these factors are not so small when compared to the
isospin-breaking factors $\delta_{K^{(\ast)}}/\Lambda_{D_{s0}^\ast}^2$
arising from the mass differences of kaons $K^{(\ast)}$.
We conclude from our results, that the heavy quark limit is not
a good approximation for the isospin-violating strong decay
$D_{s0}^{\ast} \to D_s \pi$, since some of the important isospin-breaking
effects are missing. In addition, taking in general the HQL in the charm
sector is not necessarily a good approximation because of the relatively
small mass of the charm quark. In contrast we show below that for the
radiative decay $D_{s0}^{\ast} \to D_s^{\ast} \gamma$ the HQL works well.

In Table 2 we present our results for the decay width
$\Gamma(D_{s0}^{\ast} \to D_s \pi)$ including a variation of the scale
parameter $\Lambda_{D_{s0}^\ast}$ from 1 to 2 GeV (increase of
$\Lambda_{D_{s0}^\ast}$ leads to an increase of the width) and compare
them to previous theoretical predictions.

Now we turn to the discussion of the radiative decay
$D_{s0}^{\ast} \to D_s^{\ast} \gamma$. By construction, using
a gauge-invariant and Lorentz-covariant effective Lagrangian, the full
amplitude for this process is gauge-invariant, while the separate
contributions of the different diagrams of Fig.4 are not.
It is important to stress that the diagrams of Fig.4  fall into
two separately gauge-invariant sets: one set includes the diagrams
of Figs.4(a), 4(c), 4(e), and 4(g) (with loops containing virtual 
$D^+$ and $K^0$ mesons), generated by the coupling of $D_{s0}^\ast$ to the
$D^+$ and $K^0$ constituents. Another set contains the diagrams
of Figs.4(b), 4(d), 4(f), and 4(h) (with loops containing virtual $D^0$ 
and $K^+$ mesons) with the coupling of $D_{s0}^\ast$ to $D^0$ and $K^+$. 

For convenience we split each individual diagram into a
gauge-invariant piece and a reminder, which is non-invariant.
One can prove that the sum of the non-invariant terms vanishes
due to gauge invariance. In the following discussion of the numerical
results we will deal only with the gauge-invariant contribution of 
the separate diagrams of Fig.4. Another important feature of the
$D_{s0}^{\ast} \to D_s^{\ast} \gamma$ amplitude is that the effective
coupling $G_{D_{s0}^{\ast} D_s^{\ast} \gamma}$ is ultraviolet (UV) finite.
In the Appendix we discuss the local limit that is when we remove the cutoff
with $\Lambda_{D_{s0}^\ast} \to \infty$ in the correlation function
$\Phi_{D_{s0}^\ast}$. Again, the separate contributions of the diagrams
of Fig.4 to $G_{D_{s0}^{\ast} D_s^{\ast} \gamma}$ contain divergences
which compensate each other. In the Appendix we discuss this issue in detail.

First, we show the results for the effective coupling constant
$G_{D_{s0}^{\ast} D_s^{\ast} \gamma}$: the total result and partial
contributions of the different diagrams of Fig.4
(marked by 4(a), 4(b), etc.).
In the analysis of the electromagnetic
decay $D_{s0}^{\ast} \to D_s^\ast \gamma$ we restrict to the isospin limit,
i.e. we do not include the isospin-breaking effects in the meson masses and
proceed with the masses of the charged particles. In the isospin limit
the diagrams of Fig.4(e) and 4(f) compensate each other (and therefore do not
contribute to the total amplitude), while
the diagrams of Fig.4(g) and 4(h) are equal to each other.
For a value of $\Lambda_{D_{s0}^\ast} = 1$ GeV we get
\eq
& &G_{D_{s0}^{\ast} D_s^\ast \gamma} = 0.093 \ {\rm GeV}^{-1} \,,
\hspace*{.3cm}
G_{D_{s0}^{\ast} D_s^\ast \gamma}^{4a} =
-0.030 \ {\rm GeV}^{-1} \,,
\hspace*{.3cm}
G_{D_{s0}^{\ast} D_s^\ast \gamma}^{4b} =
0.089 \ {\rm GeV}^{-1} \,, \\
& &G_{D_{s0}^{\ast} D_s^\ast \gamma}^{4c} =
10^{-4} \ {\rm GeV}^{-1} \,,
\hspace*{.4cm}
G_{D_{s0}^{\ast} D_s^\ast \gamma}^{4d} =
0.002 \ {\rm GeV}^{-1} \,,
\hspace*{.7cm}
G_{D_{s0}^{\ast} D_s^\ast \gamma}^{4g} \equiv
G_{D_{s0}^{\ast} D_s^\ast \gamma}^{4h} =
0.016 \ {\rm GeV}^{-1} \,. \nonumber
\en
The corresponding results for the decay width
$D_{s0}^{\ast} \to D_s^\ast \gamma$ are:
\eq
& &\Gamma(D_{s0}^\ast \to D_s^\ast \gamma) = 0.47 \ {\rm KeV}\,, \\
& &\Gamma(D_{s0}^\ast \to D_s^\ast \gamma)^{\rm 4a} = 0.05 \ {\rm KeV}\,,
\hspace*{.25cm}
\Gamma(D_{s0}^\ast \to D_s^\ast \gamma)^{\rm 4b} = 0.43 \ {\rm KeV}\,, 
\nonumber\\
& &\Gamma(D_{s0}^\ast \to D_s^\ast \gamma)^{\rm 4c} =
6 \times 10^{-7} \ {\rm KeV}\,,
\hspace*{.25cm}
\Gamma(D_{s0}^\ast \to D_s^\ast \gamma)^{\rm 4d} =
2 \times 10^{-4} \ {\rm KeV}\,, \nonumber\\
& &\Gamma(D_{s0}^\ast \to D_s^\ast \gamma)^{\rm 4g} \equiv
   \Gamma(D_{s0}^\ast \to D_s^\ast \gamma)^{\rm 4h} \equiv
   0.02 \ {\rm KeV}\,. \nonumber 
\en
From the results it is clear that the contact diagrams of Fig.4(c)  
and 4(d) are strongly suppressed, these diagrams are kept to guarantee 
gauge invariance. The main contribution comes from the diagram of 
Fig.4(b) where the photon couples to the $K^+$. The diagram of Fig.4(a)  
is relatively suppressed as $\sim (m_K/m_D)^2$.

The sum of all the diagrams is ultraviolet finite and, therefore, the  
cutoff parameter can be removed with $\Lambda_{D_{s0}^\ast} \to \infty$. 
In the local approximation
(LC case) for the radiative decay width we get the following
results for the coupling constant $G_{D_{s0}^{\ast} D_s^\ast \gamma}$
and the decay width $\Gamma(D_{s0}^\ast \to D_s^\ast \gamma)$ [Here
we only deal with the gauge-invariant parts of diagrams of 
Figs.4(a), 4(b), 4(g), and 4(h).]:   
\eq
& &G_{D_{s0}^{\ast} D_s^\ast \gamma} = 0.110 \ {\rm GeV}^{-1} \,,
\nonumber\\
& &G_{D_{s0}^{\ast} D_s^\ast \gamma}^{4a} = -0.038\ {\rm GeV}^{-1} \,,
\hspace*{.5cm}
G_{D_{s0}^{\ast} D_s^\ast \gamma}^{4b} = 0.093\ {\rm GeV}^{-1} \,,
\nonumber\\
& &G_{D_{s0}^{\ast} D_s^\ast \gamma}^{4g} \equiv
G_{D_{s0}^{\ast} D_s^\ast \gamma}^{4h} = 0.055 \ {\rm GeV}^{-1} \,,
\en
and
\eq
& &\Gamma(D_{s0}^\ast \to D_s^\ast \gamma) = 0.66 \ {\rm KeV}\,,
\nonumber\\
& &\Gamma(D_{s0}^\ast \to D_s^\ast \gamma)^{\rm 4a} = 0.08 \ {\rm KeV}\,,
\hspace*{.25cm}
\Gamma(D_{s0}^\ast \to D_s^\ast \gamma)^{\rm 4b} = 0.47 \ {\rm KeV}\,,
\nonumber\\
& &\Gamma(D_{s0}^\ast \to D_s^\ast \gamma)^{\rm 4g} \equiv
   \Gamma(D_{s0}^\ast \to D_s^\ast \gamma)^{\rm 4h} \equiv
   0.04 \ {\rm KeV}\,.
\en
The LC results are larger than for the nonlocal case (NC case) choosing
$\Lambda_{D_{s0}^\ast} = 1$ GeV.

Finally, we consider the HQL to this process.
In the NCHQL case the diagrams of Fig.4 relatively scale as:
\eq
G_{D_{s0}^{\ast} D_s^\ast \gamma}^{4a} \ : \
G_{D_{s0}^{\ast} D_s^\ast \gamma}^{4b} \ : \
G_{D_{s0}^{\ast} D_s^\ast \gamma}^{4c} \ : \
G_{D_{s0}^{\ast} D_s^\ast \gamma}^{4d} \ : \
G_{D_{s0}^{\ast} D_s^\ast \gamma}^{4g(h)} \ = \
\frac{1}{m_c} \ : \ 1 \ : \
\frac{1}{m_c^2} \ : \ 1 \ : \ \frac{1}{m_c} \,.
\en
Therefore, the leading order contribution arises from the diagrams of
Fig.4(b) and 4(d), resulting in
\eq
& &G_{D_{s0}^{\ast} D_s^\ast \gamma} = 0.114 \ {\rm GeV}^{-1} \,,
\nonumber\\
& &
G_{D_{s0}^{\ast} D_s^\ast \gamma}^{4b} = 0.053\ {\rm GeV}^{-1} \,,
\hspace*{.5cm}
G_{D_{s0}^{\ast} D_s^\ast \gamma}^{4d} = 0.061 \ {\rm GeV}^{-1} \,,
\en
and the corresponding results for the decay width of
\eq
& &\Gamma(D_{s0}^\ast \to D_s^\ast \gamma) = 0.71 \ {\rm KeV}\,,
\nonumber\\
& &\Gamma(D_{s0}^\ast \to D_s^\ast \gamma)^{\rm 4b} = 0.15 \ {\rm KeV}\,,
\hspace*{.25cm}
\Gamma(D_{s0}^\ast \to D_s^\ast \gamma)^{\rm 4d} = 0.20 \ {\rm KeV}\,.
\en
Finally, in the LCHQL case the diagrams of Fig.4
relatively scale as:
\eq
G_{D_{s0}^{\ast} D_s^\ast \gamma}^{4a} \ : \
G_{D_{s0}^{\ast} D_s^\ast \gamma}^{4b} \ : \
G_{D_{s0}^{\ast} D_s^\ast \gamma}^{4g(h)} \ = \
{\rm ln}\frac{m_c}{m_K} \ : \ 1 \ : \ 1
\en
Therefore, the leading order contribution comes from the diagram of
Fig.4(b) with
\eq
G_{D_{s0}^{\ast} D_s^\ast \gamma} =
\frac{g_{_{D_{s0}^\ast}} g_{_{D_s^\ast DK}}}{(4 \pi m_c)^2} \,
{\rm ln}\frac{m_c^2}{m_K^2} = 0.160 \ {\rm GeV}^{-1} \,,
\en
and
\eq
\Gamma(D_{s0}^\ast \to D_s^\ast \gamma) = 1.41 \ {\rm KeV}\,,
\en
where the coupling $g_{_{D_{s0}^\ast}}$ is given by Eq.~(\ref{HQL_loc}).

In Table 3 we summarize our results for 
$\Gamma(D_{s0}^{\ast} \to D_s^\ast \gamma)$ for all four cases
(NC, LC, NCHQL and LCHQL) including a variation of the scale
parameter $\Lambda_{D_{s0}^\ast}$ from 1 to 2 GeV (an increase of
$\Lambda_{D_{s0}^\ast}$ leads to a larger value for the width).
We also compare to predictions of other theoretical approaches.
Our results have a negligible dependence on the parameter
$\Lambda_{D_{s0}^\ast}$ and are also in good agreement
with previous calculations.
Also, within a factor of 2 our results for the different considered
cases are in good agreement. Hence for the radiative decay
$D_{s0}^{\ast} \to D_s^{\ast} \gamma$ the local approximation (LC) 
and HQL are reasonable approximations.

\section{Summary}

We studied the new charm-strange meson $D_{s0}^\ast(2317)$
in the hadronic molecule interpretation, considering a bound state
of $D$ and $K$ mesons. Using an effective Lagrangian approach
we calculated the strong $D_{s0}^{\ast} \to D_s \pi^0$ and
radiative $D_{s0}^{\ast} \to D_s^{\ast} \gamma$ decays.
A new impact of the $DK$ molecular structure
of the $D_{s0}^\ast(2317)$ meson is that the presence of $u(d)$ quarks in
the $D$ and $K$ meson loops gives rise to a direct strong isospin-violating
transition $D_{s0}^{\ast} \to D_s \pi^0$ in addition to
the decay mechanism induced by $\eta-\pi^0$ mixing as was considered
before in the literature. We showed that the direct transition dominates
over the $\eta-\pi^0$ mixing transition. Our results for the partial
decay widths are summarized as follows:
\eq
& &\Gamma(D_{s0}^\ast \to D_s \pi) = 79.3 \pm 32.6
\ {\rm KeV}  \ \ \ [{\rm "Full" \ calculation}]\,, \nonumber\\
& &\Gamma(D_{s0}^\ast \to D_s \pi) = 79.6 \pm 33.0
\ {\rm KeV}  \ \ \ [{\rm "LO" \ calculation}]\,,\nonumber\\
& &\\
& &\Gamma(D_{s0}^\ast \to D_s^\ast \gamma) = 0.55 \pm 0.08 \ {\rm KeV}
 \ \ \ [{\rm "NC" \ case}]\,,\nonumber\\
& &\Gamma(D_{s0}^\ast \to D_s^\ast \gamma) = 0.66 \ {\rm KeV}
 \hspace*{1.4cm} [{\rm "LC" \ case}]\,,\nonumber\\
& &\Gamma(D_{s0}^\ast \to D_s^\ast \gamma) = 0.94 \pm 0.23 \ {\rm KeV}
 \ \ \ [{\rm "NCHQL" \ case}]\,,\nonumber\\
& &\Gamma(D_{s0}^\ast \to D_s^\ast \gamma) = 1.41 \ {\rm KeV}
  \hspace*{1.4cm}  [{\rm "LCHQL" \ case}]\,.\nonumber
\en
The ratio
$R = \Gamma(D_{s0}^\ast \to D_s^\ast \gamma/\Gamma(D_{s0}^\ast \to D_s \pi)
\sim 10^{-2}$ satisfies the current experimental upper 
limit of $R < 0.059$~\cite{Yao:2006px}.

For the case of the strong decay the application of the
heavy quark limit (HQL) gives a significant suppression of
the direct mode. The contributions of the isospin-breaking
effects associated with the mass-difference of
$D^{(\ast)}$ mesons have an extra factor $\Lambda_{D_{s0}^\ast}/m_c$ and,
therefore, are formally of higher-order in the $1/m_c$ expansion in
comparison to the isospin-breaking effects associated with
the mass difference of  $K^{(\ast)}$ mesons. However, numerically the
factor $\Lambda_{D_{s0}^\ast}/m_c$ is of order 1, leading to the result
that the HQL is not a suitable approximation for the isospin-violating
decay $D_{s0}^{\ast} \to D_s \pi^0$.

In the case of the radiative decay $D_{s0}^{\ast} \to D_s^{\ast} \gamma$
we have another situation and the different limiting cases (local limit,
heavy quark limit) considered give more or less a similar description
of the physical quantities $G_{D_{s0}^{\ast} D_s^\ast \gamma}$ and
$\Gamma(D_{s0}^\ast \to D_s^\ast \gamma)$ (see the results of Table 3).
Here our conclusion is that in the context of a molecular interpretation
the decay width $\Gamma(D_{s0}^\ast \to D_s^\ast \gamma)$ is of order
1 KeV as was previously predicted before by other theoretical approaches.

\begin{acknowledgments}

This work was supported by the DFG under contracts FA67/31-1 and
GRK683. This research is also part of the EU Integrated
Infrastructure Initiative Hadronphysics project under contract
number RII3-CT-2004-506078 and President grant of Russia
"Scientific Schools"  No. 5103.2006.2.

\end{acknowledgments}

\newpage

\appendix\section{Matrix element of the radiative decay
$D_{s0}^{\ast} \to D_s^\ast \gamma$}

Here we discuss the matrix element of the radiative decay
$D_{s0}^{\ast} \to D_s^\ast \gamma$ in the local approximation
(when the cutoff in the $D_{s0}^\ast$ meson correlation function
is removed with $\Lambda_{D_{s0}^\ast} \to \infty$)
and for the nonlocal case.

As we mentioned before the on-shell matrix element describing
the $D_{s0}^{\ast} \to D_s^{\ast} \gamma$ transition
can be written in the manifestly gauge-invariant form
\eq
M_{\mu\nu}(D_{s0}^{\ast} \to D_s^{\ast} \gamma) \, = \,
e \, G_{D_{s0}^{\ast} D_s^{\ast} \gamma} \,
( g_{\mu\nu} p^\prime q - p^\prime_\mu q_\nu ) \,.
\en
In the local approximation the following diagrams of Fig.4
contribute to this matrix element: diagrams of Figs.4(a), 
4(b), and 4(e)-4(h). 
As we stressed in Sec.III, two sets of all the diagrams are
separately gauge-invariant: the set of Figs.4(a), 4(e), and 4(g) and
set of diagrams related to Figs.4(b), 4(f), and 4(h). 
For illustration we take one set [Figs.4(a), 4(e), and 4(g)] 
and prove gauge-invariance by using dimensional regularization (DR) 
for the separation of the divergent pieces which finally cancel each 
other. 

The structure integrals (we drop the occurring coupling constants)
corresponding to the diagrams of Figs.4(a), 4(e), and 4(g) are given by:

In diagram Fig.4(a)
\eq
T_{\mu\nu}^{4a} \ = \ - \int \frac{d^Dk}{(2\pi)^D i} \,
\frac{(2 k + p + p^\prime)_\mu \, (2k+ p^\prime)_\nu}
{[m_D^2 - (k+p)^2] \, [m_D^2 - (k+p^\prime)^2] \, [m_K^2 - k^2]} \,,
\en

In diagram Fig.4(e)
\eq
T_{\mu\nu}^{4e} \ = \ - g_{\mu\nu} \int \frac{d^Dk}{(2\pi)^D i} \, 
\frac{1}{[m_D^2 - (k+p)^2] \, [m_K^2 - k^2]}
\en

In diagram Fig.4(g)
\eq
T_{\mu\nu}^{4g} \ = \ \Gamma_{\mu\nu\alpha} \,\,
\frac{- g^{\alpha\beta} + p^\alpha p^\beta / m_{D_s^\ast}^2}
{m_{D_s^\ast}^2 - p^2} \, \, \int \frac{d^Dk}{(2\pi)^D i} \,
\frac{(2k+ p^\prime)_\beta}
{[m_D^2 - (k+p)^2] \, [m_D^2 - (k+p^\prime)^2] \, [m_K^2 - k^2]}\,,
\en
where
\eq
\Gamma_{\mu\nu\alpha} \, = \,
\, - \, g_{\nu\alpha} \, (p + p^\prime)_\mu
\, + \, \frac{g_{\mu\alpha}}{2} \, (p + p^\prime)_\nu \,
\, + \, \frac{g_{\mu\nu}}{2}    \, (p + p^\prime)_\alpha \, .
\en
Next using the Feynman $\alpha$-parametrization and
the master formula of DR
\eq
\int \frac{d^Dk}{(2\pi)^D i} \,
\frac{(-k^2)^M}{[\Delta - k^2]^N} =
\frac{1}{(4\pi)^{D/2}} \, \frac{\Gamma(D/2+M) \, \Gamma(N-M-D/2)}
{\Gamma(D/2) \, \Gamma(N)} \, \Delta^{D/2+M-N}
\en
we get:
\eq\label{T4a}
T_{\mu\nu}^{4a} &=&
\frac{g_{\mu\nu}}{16 \pi^2} \biggl\{ \frac{2}{4-D} +
{\rm ln}4\pi + \Gamma^\prime(1) \biggr\}
- \frac{g_{\mu\nu}}{8 \pi^2}
\int\limits_0^1 d\alpha (1-\alpha) {\rm ln}\Delta_{DK} \nonumber\\
&-& \frac{1}{4\pi^2} (g_{\mu\nu} p^\prime q - p^\prime_\mu q_\nu) \,
\int\limits_0^1 d^3\alpha \,
\delta\biggl(1 - \sum\limits_{i=1}^{3} \, \alpha_i\biggr) \,
\frac{\alpha_1 \alpha_3}{\Delta_{DDK}} + O(D-4) \,,
\en
\eq\label{T4e}
T_{\mu\nu}^{4e} &=& - \frac{g_{\mu\nu}}{16 \pi^2}
\biggl\{ \frac{2}{4-D} +
{\rm ln}4\pi + \Gamma^\prime(1) \biggr\}
+ \frac{g_{\mu\nu}}{16 \pi^2}
\int\limits_0^1 d\alpha \, {\rm ln}\Delta_{DK} + O(D-4) \,,
\en
\eq\label{T4g}
T_{\mu\nu}^{4g} \ = \  \frac{g_{\mu\nu}}{16 \pi^2}
\int\limits_0^1 d\alpha (1 - 2\alpha) \, {\rm ln}\Delta_{DK}
\, + \, \frac{3}{32 \pi^2 m_{D_s^\ast}^2}
\, (g_{\mu\nu} p^\prime q - p^\prime_\mu q_\nu) \,
\int\limits_0^1 d\alpha (1 - 2 \alpha) {\rm ln}\Delta_{DK} + O(D-4) \,,
\en
where
\eq
\Delta_{DDK} &=& \Delta_3(m_D,m_K) =
m_D^2 (1 - \alpha_3) + m_K^2 \alpha_3
- m_{D_{s0}^\ast}^2 \alpha_1 \alpha_3
- m_{D_{s}^\ast}^2 \alpha_2 \alpha_3 \,, \\
\Delta_{DK} &=& \Delta_2(m_D,m_K) = m_D^2 (1 - \alpha) + m_K^2 \alpha
- m_{D_{s0}^\ast}^2 \alpha (1 - \alpha)  \,. \nonumber
\en
From Eqs.~(\ref{T4a})-(\ref{T4g}) one can see that in the sum of
the diagrams of Figs.4(a), 4(e), and 4(g) all divergences and
non-gauge invariant pieces cancel each other. Taking $D \to 4$
we write down the final result of:
\eq\label{Taeg}
T_{\mu\nu}^{4a+4e+4g} \ = \
\frac{1}{4\pi^2} (g_{\mu\nu} p^\prime q - p^\prime_\mu q_\nu) \,
\biggl\{
- \int\limits_0^1 d^3\alpha \,
\delta\biggl(1 - \sum\limits_{i=1}^{3} \, \alpha_i\biggr) \,
\frac{\alpha_1 \alpha_3}{\Delta_{DDK}}
\, + \, \frac{3}{8 m_{D_s^\ast}^2} \,
\int\limits_0^1 d\alpha (1 - 2 \alpha) {\rm ln}\Delta_{DK} \biggr\}
\en
By analogy we prove the gauge invariance for the sum of the
diagrams of Figs.4(b), 4(f), and 4(h):
\eq\label{Tbfh}
T_{\mu\nu}^{4b+4f+4h} \ = \
\frac{1}{4\pi^2} (g_{\mu\nu} p^\prime q - p^\prime_\mu q_\nu) \,
\biggl\{
\int\limits_0^1 d^3\alpha \,
\delta\biggl(1 - \sum\limits_{i=1}^{3} \, \alpha_i\biggr) \,
\frac{\alpha_1 \alpha_3}{\Delta_{KKD}}
\, - \, \frac{3}{8 m_{D_s^\ast}^2} \,
\int\limits_0^1 d\alpha (1 - 2 \alpha) {\rm ln}\Delta_{KD} \biggr\}
\en
where
$\Delta_{KKD} = \Delta_3(m_K,m_D)$ and
$\Delta_{KD}  = \Delta_2(m_K,m_D)$.
It is easy to show that the second terms in
Eqs.~(\ref{Taeg}) and (\ref{Tbfh}) are equal to each other
by changing the variable $\alpha$ to $1 - \alpha$.
Therefore, the total result for the effective coupling constant
$G_{D_{s0}^{\ast} D_s^{\ast} \gamma}$ in the local case is:
\eq
G_{D_{s0}^{\ast} D_s^{\ast} \gamma} =
\frac{g_{_{D_{s0}^\ast}} g_{_{D_s^\ast DK}}}{4 \pi^2} \,
\biggl\{
\int\limits_0^1 d^3\alpha \,
\delta\biggl(1 - \sum\limits_{i=1}^{3} \, \alpha_i\biggr) \,
\alpha_1 \alpha_3 \biggl\{ \frac{1}{\Delta_{KKD}} -
\frac{1}{\Delta_{DDK}} \biggr\}
\, + \, \frac{3}{4 m_{D_s^\ast}^2} \,
\int\limits_0^1 d\alpha (1 - 2 \alpha) {\rm ln}\Delta_{DK} \biggr\} \,.
\en
In the nonlocal case the gauge invariance can be proved
based on a method developed e.g. in Ref.~\cite{Faessler:2003yf}.
For this purpose in particular we split the contribution of each diagram 
into a part which is gauge invariant and one which is not:  
we use the following representation for the four-vectors 
with open Lorentz indices $\mu$ and $\nu$: 
\begin{eqnarray}\label{split}
p^\mu \, = \, p^\mu_{\perp; \, q} \,
+ \, q^\mu \, \frac{p q}{q^2}\,, \\
p^\nu \, = \, p^\nu_{\perp; \, p^\prime} \,
+ \, p^{\prime\nu} \, \frac{p p^\prime}{p^{\prime 2}}\,,
\nonumber
\end{eqnarray}
such that $p^\mu_{\perp; \, q} \, q_\mu=0$ and
$p^\nu_{\perp; \, p^\prime} \, p^\prime_\nu=0$.
Expressions for diagrams containing only $\perp$-values are gauge invariant
separately.
It is easy to show that the remaining terms, which are not gauge invariant,
cancel each other in total. Note that this method works perfectly both
for on-shell and off-shell amplitudes.

The coupling constant $G_{D_{s0}^{\ast} D_s^{\ast} \gamma}$ in the
nonlocal case is given by
\eq
G_{D_{s0}^{\ast} D_s^{\ast} \gamma} =
\frac{g_{_{D_{s0}^\ast}} g_{_{D_s^\ast DK}}}{16 \pi^2} \,
I_{D_{s0}^{\ast} D_s^{\ast} \gamma} \,,
\en
where $I_{D_{s0}^{\ast} D_s^{\ast} \gamma}$ is the structure integral
containing the contributions of the diagrams in 
Figs.4(a)-4(d), 4(g), and 4(h):
\eq
I_{D_{s0}^{\ast} D_s^{\ast} \gamma} &=&
\sum\limits_{i=a,b,c,d,g,h} I_{D_{s0}^{\ast} D_s^{\ast} \gamma}^{4i}
\,,\nonumber\\
I_{D_{s0}^{\ast} D_s^{\ast} \gamma}^{4a} &=&
- \frac{4}{\Lambda^2} \int\limits_0^\infty
\int\limits_0^\infty \int\limits_0^\infty \,
\frac{d\alpha_1d\alpha_2d\alpha_3}{(1+\alpha_{123})^4} \,
(\alpha_1 + w_K) (\alpha_3 + w_D) \,\,
[ - d\tilde \Phi^\prime_{D_{s0}^\ast}(z_{DDK})] \,, \nonumber\\
I_{D_{s0}^{\ast} D_s^{\ast} \gamma}^{4b} &=&
 \frac{4}{\Lambda^2} \int\limits_0^\infty
\int\limits_0^\infty \int\limits_0^\infty \,
\frac{d\alpha_1d\alpha_2d\alpha_3}{(1+\alpha_{123})^4} \,
(\alpha_1 + w_D) (\alpha_3 + w_K) \,\,
[ - d\tilde \Phi^\prime_{D_{s0}^\ast}(z_{KKD})] \,, \\
I_{D_{s0}^{\ast} D_s^{\ast} \gamma}^{4c} &=&
 \frac{4}{\Lambda^2} w_K^2 \int\limits_0^1 dt t
\int\limits_0^\infty\int\limits_0^\infty
\frac{d\alpha_1d\alpha_2}{(1+\alpha_{12})^4} \,
(w_D \alpha_1 - w_K \alpha_2) \,\,
[ - d\tilde \Phi^\prime_{D_{s0}^\ast}(z_{DK})] \,, \nonumber\\
I_{D_{s0}^{\ast} D_s^{\ast} \gamma}^{4d} &=&
 \frac{4}{\Lambda^2} w_D^2 \int\limits_0^1 dt t
\int\limits_0^\infty\int\limits_0^\infty
\frac{d\alpha_1d\alpha_2}{(1+\alpha_{12})^4} \,
(w_D \alpha_2 - w_K \alpha_1) \,\,
[ - d\tilde \Phi^\prime_{D_{s0}^\ast}(z_{KD})] \,, \nonumber\\
I_{D_{s0}^{\ast} D_s^{\ast} \gamma}^{4g}
&\equiv& I_{D_{s0}^{\ast} D_s^{\ast} \gamma}^{4h} \, = \, 
 \frac{3}{2 m_{D_{s0}^\ast}^2}
\int\limits_0^\infty\int\limits_0^\infty
\frac{d\alpha_1d\alpha_2}{(1+\alpha_{12})^3} \,
(\alpha_2 - \alpha_1 + w_D - w_K) \,\,
\tilde \Phi_{D_{s0}^\ast}(z_P) \,, \nonumber 
\en
where
\eq
& &\alpha_{123} = \alpha_1 + \alpha_2 + \alpha_3\,, \hspace*{.5cm}
\alpha_{12} = \alpha_1 + \alpha_2\,, \nonumber\\[2mm]
& &z_{DDK} = z_3(\mu_D,\mu_K) \,, \hspace*{.5cm}
   z_{KKD} = z_3(\mu_K,\mu_D) \,, \hspace*{.5cm}
   z_{DK}  = z_2(\mu_D,\mu_K) \,, \hspace*{.5cm}
   z_{KD}  = z_2(\mu_K,\mu_D) \,, \nonumber\\[2mm]
& &z_3(\mu_1,\mu_2) = 
\mu_1^2 \alpha_{12} + \mu_2^2 \alpha_3 + \mu_{D_{s0}^\ast}^2 w_1 w_2
- \frac{\alpha_3+w_1}{1 + \alpha_{123}} ( \mu_{D_{s0}^\ast}^2 (\alpha_1+w_2)
+ \mu_{D_{s}^\ast}^2 \alpha_2 )  \,, \\[2mm]
& &z_2(\mu_1,\mu_2) = 
\mu_1^2 \alpha_1 + \mu_2^2 \alpha_2 + ( \mu_{D_{s0}^\ast}^2 t +
\mu_{D_{s}^\ast}^2 (1 - t) ) w_1 w_2
- \frac{\alpha_2+w_1}{1 + \alpha_{12}} ( \mu_{D_{s0}^\ast}^2 w_2 t
+ \mu_{D_{s}^\ast}^2 ( w_2 (1-t) + \alpha_1 )) \,,\nonumber\\[2mm]
& &z_{P} = 
\mu_D^2 \alpha_1 + \mu_K^2 \alpha_2 + \mu_{D_{s0}^\ast}^2 
\biggl( w_D w_K - \frac{(\alpha_1+w_K)(\alpha_2+w_D)}{1 + \alpha_{12}}
\biggr) \,, \hspace*{.5cm} \mu_M = \frac{m_M}{\Lambda_{D_{s0}^\ast}} 
\,.\nonumber
\en

\vspace*{.5cm}

\begin{figure}

\vspace*{2cm}

\begin{center}
\epsfig{file=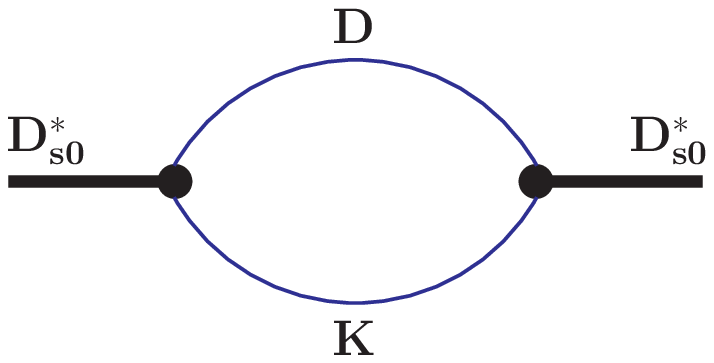, scale=.5}
\end{center}
\caption{Mass operator of the $D_{s0}^{\ast}(2317)$ meson.}

\vspace*{.2cm}

\begin{center}
\epsfig{file=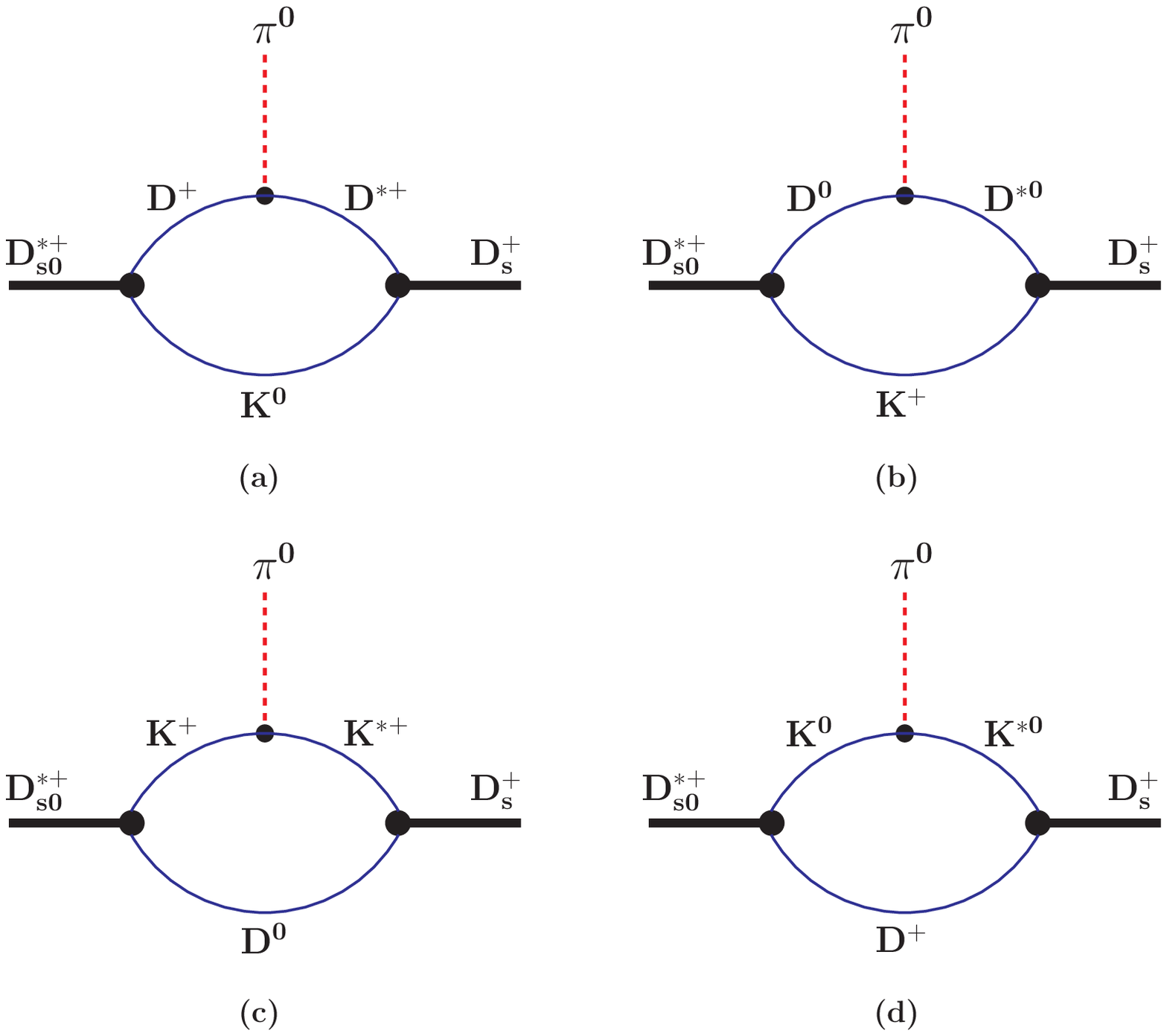, scale=.5}
\end{center}
\caption{Diagrams contributing to the ``direct'' strong
transition
$D_{s0}^{\ast +} \to D_{s}^{+} + \pi^0$.}

\vspace*{.2cm}

\begin{center}
\epsfig{file=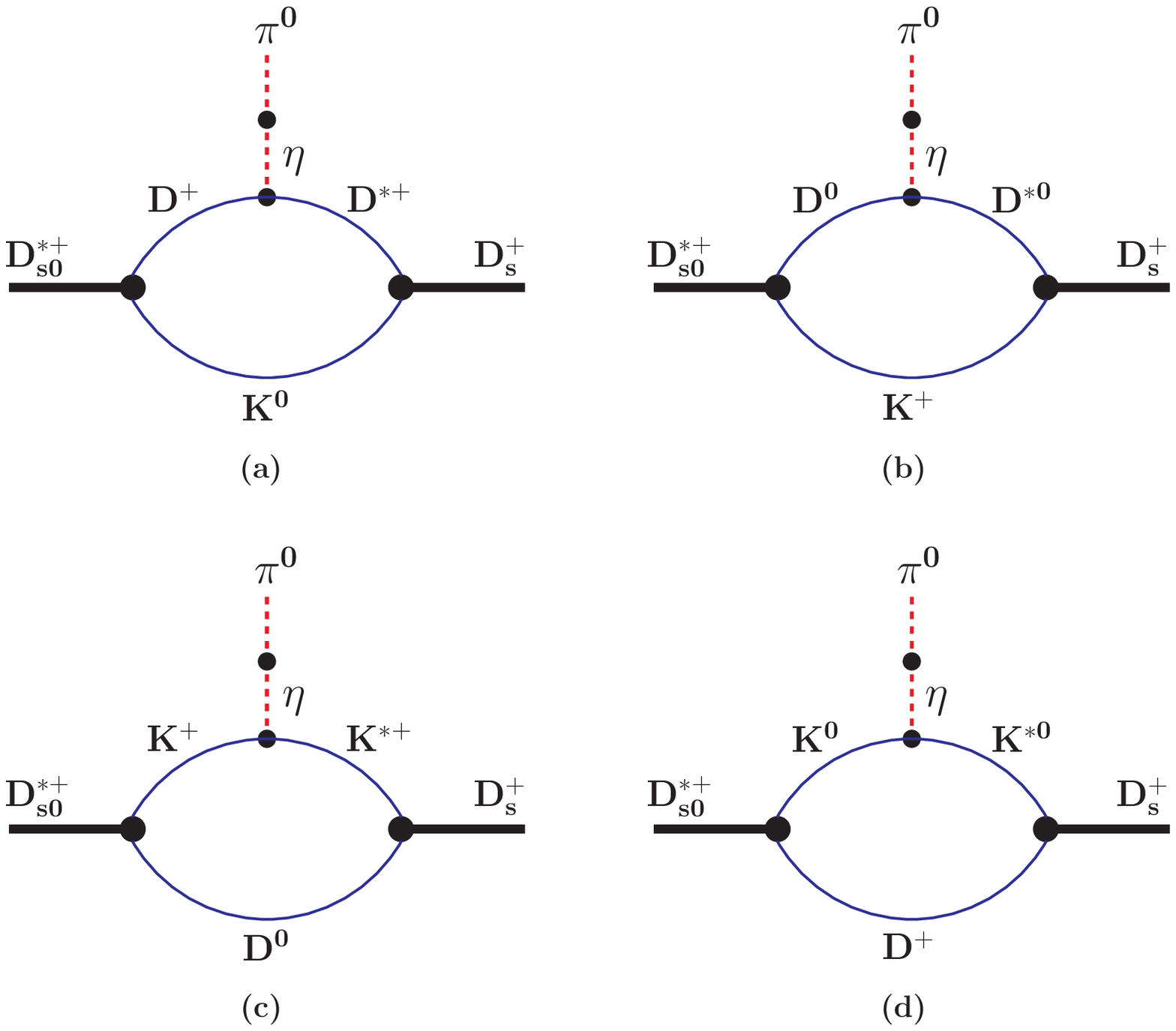, scale=.5}
\end{center}
\caption{Diagrams contributing to the strong transition
$D_{s0}^{\ast +} \to D_{s}^{+} + \pi^0$ via $\eta-\pi^0$ mixing.}
\end{figure}

\vspace*{.25cm}

\begin{figure}
\begin{center}
\epsfig{file=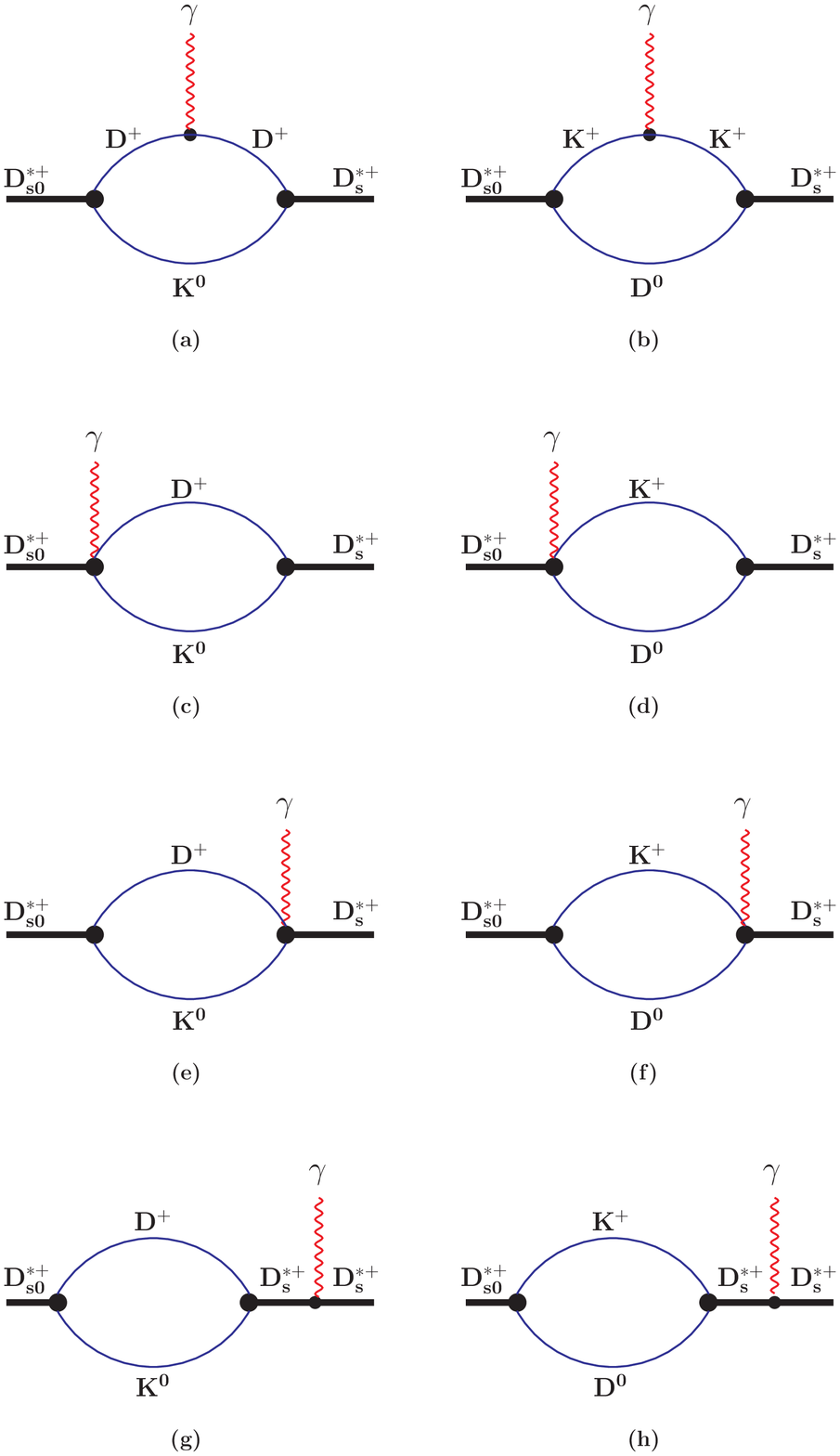, scale=.5}
\end{center}
\caption{Diagrams contributing to the radiative transition
$D_{s0}^{\ast +} \to D_{s}^{\ast +} + \gamma$.}
\end{figure}

\newpage

\begin{table}
\vspace*{2cm}
\begin{center}
{\bf Table 1.}
Coupling constant $g_{_{D_{s0}^\ast DK}}$. \\
The range of values for our results is due \\ to the 
variation of $\Lambda_{D_{s0}^\ast}$ from 1 to 2 GeV.

\vspace*{.25cm}

\def\arraystretch{1.2}
\begin{tabular}{|l|l|}
\hline
\hspace*{.5cm}
Approach \hspace*{.5cm}
& \hspace*{.5cm} $g_{_{D_{s0}^\ast DK}}$ (GeV) \hspace*{.5cm} \\
\hline
\,\,\,\,\,
Ref.~\cite{Nielsen:2006mi}
\,\,\,\,\, & \,\,\,\,\,\,\,\, 2.5 - 3.8 \,\,\,\,\, \\
\,\,\,\,\,
Ref.~\cite{Becirevic:1999fr}
\,\,\,\,\, & \,\,\,\,\,\,\,\, 5.068 \,\,\,\,\, \\
\,\,\,\,\,
Ref.~\cite{Colangelo:1995ph}
& \,\,\,\,\,\,\,\, 5.5 $\pm$ 1.8 \,\,\,\,\, \\
\,\,\,\,\,
Ref.~\cite{Wang:2006id}
\,\,\,\,\, & \,\,\,\,\,\,\,\, 5.9$^{+1.7}_{-1.6}$ \,\,\,\,\, \\
\,\,\,\,\,
Ref.~\cite{Mehen:2004uj}
& \,\,\,\,\,\,\,\,  6.0 - 7.8 \,\,\,\,\, \\
\,\,\,\,\,
Ref.~\cite{Wang:2006bs}
\,\,\,\,\, & \,\,\,\,\,\,\,\, 9.3$^{+2.7}_{-2.1}$ \,\,\,\,\, \\
\,\,\,\,\,
Ref.~\cite{Chow:1995ca}
\,\,\,\,\,       & \,\,\,\,\,\,\,\, $<$ 9.86 \,\,\,\,\, \\
\,\,\,\,\,
Ref.~\cite{Guo:2006fu}
\,\,\,\,\, & \,\,\,\,\,\,\,\, 10.203 \,\,\,\,\, \\
\hline
\,\,\,\,\, Our results:                 & \\
\,\,\,\,\, NC case
& \,\,\,\,\,\,\,\, 9.90 $-$ 11.26 \,\,\,\,\, \\
\,\,\,\,\, LC case
& \,\,\,\,\,\,\,\, 8.98 \,\,\,\,\, \\
\,\,\,\,\, NCHQL case
& \,\,\,\,\,\,\,\, 11.52 $-$ 16.22 \,\,\,\,\, \\
\,\,\,\,\, LCHQL case
& \,\,\,\,\,\,\,\, 11.52 \,\,\,\,\, \\
\hline
\end{tabular}
\end{center}

\vspace*{1cm}

\begin{center}
{\bf Table 2.}
Decay width of $D_{s0}^{\ast} \to D_s \pi^0$. \\ 
The range of values for our results is due \\ to the 
variation of $\Lambda_{D_{s0}^\ast}$ from 1 to 2 GeV.

\vspace*{.25cm}

\def\arraystretch{1.2}
\begin{tabular}{|l|l|}
\hline
\hspace*{.5cm}
Approach \hspace*{.5cm}
& \hspace*{.5cm}
$\Gamma(D_{s0}^{\ast} \to D_s \pi^0)$ (KeV) \hspace*{.5cm} \\
\hline
\,\,\,\,\,
Ref.~\cite{Nielsen:2005zr}
\,\,\,\,\, & \,\,\,\,\,\,\,\, 6 $\pm$ 2 \,\,\,\,\, \\
\,\,\,\,\,
Ref.~\cite{Colangelo:2003vg}
\,\,\,\,\, & \,\,\,\,\,\,\,\, 7 $\pm$ 1 \,\,\,\,\, \\
\,\,\,\,\,
Ref.~\cite{Godfrey:2003kg}
\,\,\,\,\, & \,\,\,\,\,\,\,\, 10 \,\,\,\,\, \\
\,\,\,\,\,
Ref.~\cite{Fayyazuddin:2003dp}
& \,\,\,\,\,\,\,\, 16 \,\,\,\,\, \\
\,\,\,\,\,
Ref.~\cite{Bardeen:2003kt}
\,\,\,\,\, & \,\,\,\,\,\,\,\, 21.5 \,\,\,\,\, \\
\,\,\,\,\,
Ref.~\cite{Lu:2006ry}
& \,\,\,\,\,\,\,\, 32 \,\,\,\,\, \\
\,\,\,\,\,
Ref.~\cite{Wei:2005ag}
\,\,\,\,\,       & \,\,\,\,\,\,\,\, 39 $\pm$ 5 \,\,\,\,\, \\
\,\,\,\,\,
Ref.~\cite{Ishida:2003gu}
\,\,\,\,\, & \,\,\,\,\,\,\,\, 15 $-$ 70 \,\,\,\,\, \\
\,\,\,\,\,
Ref.~\cite{Cheng:2003kg}
\,\,\,\,\, & \,\,\,\,\,\,\,\, 10 $-$ 100 \,\,\,\,\, \\
\,\,\,\,\,
Ref.~\cite{Azimov:2004xk}
\,\,\,\,\, & \,\,\,\,\,\,\,\, 129 $\pm$ 43 (109 $\pm$ 16) \,\,\,\,\, \\
\hline
\,\,\,\,\, Our results:                 & \\
\,\,\,\,\, Full case
& \,\,\,\,\,\,\,\, 46.7 $-$ 111.9 \,\,\,\,\, \\
\,\,\,\,\, LO case
& \,\,\,\,\,\,\,\, 46.6 $-$ 112.6 \,\,\,\,\, \\
\hline
\end{tabular}
\end{center}
\end{table}

\newpage 

\begin{table}
\begin{center}
{\bf Table 3.}
Decay width of $D_{s0}^{\ast} \to D_s^{\ast} \gamma \ $. \\ 
The range of values for our results is due \\ to the 
variation of $\Lambda_{D_{s0}^\ast}$ from 1 to 2 GeV.

\vspace*{.25cm}

\def\arraystretch{1.2}
\begin{tabular}{|l|l|}
\hline
\hspace*{.5cm}
Approach \hspace*{.5cm}
& \hspace*{.5cm}
$\Gamma(D_{s0}^{\ast} \to D_s^{\ast} \gamma) \ \ $ (KeV) \hspace*{.5cm} \\
\hline
\,\,\,\,\,
Ref.~\cite{Fayyazuddin:2003dp}
& \,\,\,\,\,\,\,\, 0.2 \,\,\,\,\, \\
\,\,\,\,\,
Ref.~\cite{Colangelo:2003vg}
\,\,\,\,\, & \,\,\,\,\,\,\,\, 0.85 $\pm$ 0.05 \,\,\,\,\, \\
\,\,\,\,\,
Ref.~\cite{Close:2005se}
\,\,\,\,\, & \,\,\,\,\,\,\,\, 1 \,\,\,\,\, \\
\,\,\,\,\,
Ref.~\cite{Liu:2006jx}
\,\,\,\,\, & \,\,\,\,\,\,\,\, 1.1 \,\,\,\,\, \\
\,\,\,\,\,
Ref.~\cite{Wang:2006mf}
\,\,\,\,\, & \,\,\,\,\,\,\,\, 1.3 $-$ 9.9 \,\,\,\,\, \\
\,\,\,\,\,
Ref.~\cite{Azimov:2004xk}
\,\,\,\,\, & \,\,\,\,\,\,\,\, $\le$ 1.4 \,\,\,\,\, \\
\,\,\,\,\,
Ref.~\cite{Bardeen:2003kt}
\,\,\,\,\, & \,\,\,\,\,\,\,\, 1.74 \,\,\,\,\, \\
\,\,\,\,\,
Ref.~\cite{Godfrey:2003kg}
\,\,\,\,\, & \,\,\,\,\,\,\,\, 1.9 \,\,\,\,\, \\
\,\,\,\,\,
Ref.~\cite{Colangelo:2005hv}
& \,\,\,\,\,\,\,\, 4 $-$ 6 \,\,\,\,\, \\
\,\,\,\,\,
Ref.~\cite{Ishida:2003gu}
\,\,\,\,\, & \,\,\,\,\,\,\,\, 21 \,\,\,\,\, \\
\hline
\,\,\,\,\, Our results:                 & \\
\,\,\,\,\, NC case
& \,\,\,\,\,\,\,\, 0.47 $-$ 0.63 \,\,\,\,\, \\
\,\,\,\,\, LC case
& \,\,\,\,\,\,\,\, 0.66 \,\,\,\,\, \\
\,\,\,\,\, NCHQL case
& \,\,\,\,\,\,\,\, 0.71 $-$ 1.17 \,\,\,\,\, \\
\,\,\,\,\, LCHQL case
& \,\,\,\,\,\,\,\, 1.41 \,\,\,\,\, \\
\hline
\end{tabular}
\end{center}
\end{table}

\end{document}